%% file: polariton.tex
\documentclass[amsmath,amssymb,aip,jcp,
citeautoscript,superscriptaddress,twocolumn,
10pt]{revtex4-1}
\usepackage[normalem]{ulem}
\usepackage{bm}
\usepackage{graphicx}
\usepackage{dcolumn}
\usepackage[colorlinks,linkcolor=red,anchorcolor=green,
citecolor=blue,breaklinks]{hyperref}

\usepackage{float}
\usepackage{mathrsfs}
\usepackage{color}
\usepackage[normalem]{ulem}

\newcommand{\br}{\mathbf{r}}
\newcommand{\bR}{\mathbf{R}}
\newcommand{\bd}{\mathbf{d}}

\newcommand{\bu}{\mathbf{u}}

\newcommand{\bA}{\mathbf{A}}

\graphicspath{{figures/}}
\begin{document}


\title{Non-Adiabatic Molecular Dynamics of Molecules in the Presence of Strong Light-Matter Interactions}

\author{Yu Zhang}
\affiliation{Physics and Chemistry of Materials, Theoretical Division, Los Alamos National Laboratory, Los Alamos, New Mexico, 87545, USA}
\email{zhy@lanl.gov}

\author{Tammie Nelson}
\affiliation{Physics and Chemistry of Materials, Theoretical Division, Los Alamos National Laboratory, Los Alamos, New Mexico, 87545, USA}

\author{Sergei Tretiak}
\affiliation{Physics and Chemistry of Materials, Theoretical Division, Los Alamos National Laboratory, Los Alamos, New Mexico, 87545, USA}
\affiliation{Center for Integrated Nanotechnologies, Los Alamos National Laboratory, Los Alamos, New Mexico 87545, USA}

\date{\today}

\begin{abstract}
When the interaction between a molecular system and confined light modes in an optical or plasmonic cavity is strong enough to overcome the dissipative process, hybrid light-matter states (polaritons) become the fundamental excitations in the system. The mixing between the light and matter characters modifies the photophysical and photochemical properties.
Especially, it was reported that these polaritons can be employed to control photochemical reactions, charge and energy transfer, and other processes. 
In addition, according to recent studies, vibrational strong coupling can be employed to resonantly enhance the thermally-activated chemical reactions. In this work, a theoretical model and an efficient numerical method for studying the dynamics of molecules strongly interacting with quantum light are developed based on non-adiabatic excited-state molecular dynamics. The methodology was employed to study the \textit{cis-trans} photoisomerization of a realistic molecule in a cavity. Numerical simulations demonstrate that the photochemical reactions can be controlled by tuning the properties of the cavity. In the calculated example, the isomerization is suppressed when polaritonic states develop a local minimum on the lower polaritonic state. Moreover, the observed reduction of isomerization is tunable via the photon energy and light-molecule coupling strength. But the fluctuation in transition dipole screens the effect of light-matter, which makes it harder to tune the photochemical properties via the coupling strength. These insights suggest quantum control of photochemical reactions is possible by specially designed photonic or plasmonic cavities. 
\end{abstract}

\maketitle

\section{Introduction}\label{seq:intro}
The interaction between light and matter is an important subject in physics, chemistry, materials, and energy science~\cite{huagbook}. The light-matter interaction in many applications is usually weak, thus it is considered to be a small perturbation. Even in these cases, light has been employed as a powerful tool to change and detect the quantum state of molecules, such as initiation of specific excited state dynamics, pump-probe time-resolved experiments, two-dimensional spectroscopy, and Raman probes~\cite{C5CS00763A,Eichberger2010,Prokhorenko1257,acc7b00369,cr6b00002,cr020683w}. This picture has been the foundation of employing light in spectroscopy, and underpins our current understanding of photophysics and photochemistry~\cite{cr6b00552}. However, when the coupling between light and molecules is large enough to compete with or overcome the dissipation or dephasing (i.e. when the coherent energy exchange between a confined light mode and a quantum emitter is faster than the decay and decoherence timescales of each part), the system enters into the strong coupling regime and Rabi splittings between resulting modes become observable~\cite{prl681132,prl3314}. In this regime, photons and molecular excitations become superimposed and the fundamental excitations of the system are \textit{polaritons} which are hybrid light-matter excitations~\cite{accounts6b00295}. In this case, the frameworks used to describe the weak coupling regime become entirely invalid, and new theory and efficient numerical methods for strong light-matter coupling are required. 

To excite polaritons, a condensed ensemble of atoms or molecules is constrained within a cavity. The latter consists of a nanostructure that can confine Electromagnetic (EM) fields at the nano- or micro-scale, where the quantized EM modes are tuned to be resonant with the atomic or molecular ensemble. In practice, a parallel arrangement of mirrors that forms a standing wave is widely used as the optical cavity~\cite{photonic7b00674}. In addition, the recent experimental development of plasmonic cavities (metallic nanostructures that support surface plasmon or localized surface plasmon, LSP~\cite{jp026731y,acsnano8b03830,Trm2014}) makes it possible to confine and to enhance EM fields in a much smaller volume, resulting in strong and ultra-strong light-matter interaction at room temperature~\cite{RN638,prl118237401,RN843}. 

The mixed light-matter states of organic or inorganic polaritons have enabled a large number of interesting applications~\cite{Sanvitto2016,RN844}, including long-range energy transfer~\cite{Coupling2017}, enhanced charge transfer~\cite{Orgiu2015}, and polariton lasing~\cite{Ramezani17,KenaCohen2010}. The pioneering experiment by the group of Thomas Ebbesen showed that strong coupling could affect the landscape of the potential energy surface (PES), which in return alters the rate of photochemical reactions~\cite{Hutchison2012}. Since the mixed light-matter states have different PESs than unmixed states, the strong light-matter interaction offers an attractive way to control chemical reactions by reshaping the PES landscape. This possibility inspired the appearance of  polaritonic chemistry aiming to manipulate chemical structure and reactions via the formation of polaritons, which has become a topic of intense experimental~\cite{C4NR01971G,nl6b00310,acsphotonics6b00482} and theoretical research~\cite{acs7b00610,jctc6b01126,photonics7b00916,Galego2016,acsphotonics7b00305,jpclett8b02032} in the past few years. In addition, recent developments have found that vibrational strong coupling (VSC) can resonantly enhance thermally-activated chemical reactions via the formation of vibrational polaritons~\cite{Joelvsc2019, Thomas615}. 

Currently, the majority of microscopic models for strong coupling in polaritonic chemistry employ few-level model systems~\cite{PRL110126801,PRB88075321,jcp4919348}  to examine their influence on the polariton properties. However, organic chromophores have much more complicated electronic structure characterized by high density of electronic and vibrational states  compared to simple few-level quantum emitters. Thus, a quantitative quantum model describing a collection of organic molecules strongly coupled to a confined light mode, should combine the concepts of both quantum electrodynamcis (QED) and quantum chemistry~\cite{PRX5041022,PRL116238301,photonics7b00680,Flick3026,Flick15285}. In addition, even though nuclear degrees of freedom were considered in many works, these studies were limited to treating vibrational modes within the harmonic oscillator approximation~\cite{jcp4919348}. However, the latter approximation usually fails in photochemistry, especially in dynamics where the Born-Oppenheimer approximation frequently breaks down. Consequently, a theory incorporating non-adiabatic effects is required for describing the strong light-matter interaction in photochemistry. While rigorous methods accounting for non-adiabatic quantum dynamics have been proposed for strong light-matter interaction, the computational complexity of quantum dynamics ultimately limits its applications to minimal systems where only a few PESs and dimensions can be treated~\cite{jpclett6b00864,jcp4941053,C6FD00095A,Kowalewski3278,jpca7b11833,jpclett9b01599}. In order to balance the computational accuracy and efficiency, mixed quantum-classical dynamics (MQC) may offer an optimal approach for modeling polaritonic chemistry. Two popular MQC methods are the mean-field Ehrenfest dynamics and trajectory surface hopping (TSH)~\cite{cr7b00577}. Both methods require the calculations of electronic structures, gradients, and non-adiabatic couplings~\cite{cr7b00577}. Even though there were previous reports of MQC for polaritonic chemistry~\cite{jctc7b00388}, the non-adiabatic couplings (NACs) between polaritonic states are yet to be accurately derived and defined. 
In addition, in the previous simulations~\cite{jctc7b00388}, transitions between states were restricted to the degenerate state case. Finally, the derivative of transition dipole, critical in the calculation of the gradients of polaritonic states, was approximated to zero order. However, the transition dipole may change dramatically during the dynamics, especially in photochemical reactions where large conformational change is expected. Hence, a formalism of calculating the gradient of transition dipole on-the-fly is desired. In this work, we formulate the MQC methods accounting for the presence of strong light-matter interaction to describe the molecular motions on polaritonic PESs, where the gradients of polaritonic states and NACs between polaritonic state are rigorously derived.

The paper is organized as follows. First, the concept of strong light-matter interaction is introduced in Sec.~\ref{sec:jc}. Second, the electronic structure in the presence of strong light-matter interaction,  gradients, and non-adiabatic couplings in the Time-Dependent Self-Consistent Field (TDSCF) method are derived in Sec.~\ref{sec:eslight} and ~\ref{sec:gradients}, respectively. Third, the formalisms of Non-Adiabatic Molecular Dynamics (NAMD) for polaritonic chemistry, including mean-field Ehrenfest dynamics and Fewest-Switches Surface Hopping (FSSH) algorithm, are introduced in Sec.~\ref{sec:namd}. Then, numerical implementation and applications are presented in Sec.~\ref{numerical}. We modeled the photoisomerizaiton of stilbene to demonstrate the utility of the methodology. Stilbene is an excellent test-bed molecule because it has been subject to many previous studies that we can use it to validate our result~\cite{Galego2016,cr00003a007}.  Finally, our findings are summarized in Sec.~\ref{summary}.

\section{Non-adiabatic molecular dynamics for polaritonic chemistry}
\label{secmethod}
\subsection{Jaynes-Cummings (JC) and Tavis-Cummings (TC) models}
\label{sec:jc}
In this section, we first introduce the basic concept and properties of a polariton by studying the simplest models with strong coupling between light and matter. The optical cavities employed for molecular systems subject to strong coupling are generally made of two highly reflective metallic or dielectric mirrors separated by a certain distance. Different shapes of mirrors are used to generate different photonic modes. In recent decades, new ways to induce strong and ultrastrong light-matter interactions in plasmonic cavities have emerged~\cite{photonic7b00674}. The field of quantum optics commonly uses two-level models to examine the polaritonic behaviors. One such widely used model, the Jaynes-Cummings (JC) model~\cite{1443594} with rotating wave approximation (RWA)~\cite{qoptics},  considers a two-level system (TLS) which interacts with a lossless cavity mode of frequency $\omega_c$. 

The JC model can be generalized straightforwardly for photochemistry, where the corresponding Hamiltonian can be written as
\begin{equation}\label{eq:hjc}
\hat{H}_{JC}=\omega_c a^\dag a + \Omega(\bR) \sigma^\dag\sigma
-\bar{g}(\bR)(a^\dag\sigma+a\sigma^\dag),
\end{equation}
$a(a^\dag)$ is the cavity photon annihilation (creation) operator, $\omega_c$ is the energy of the cavity photonic mode. $\sigma(\sigma^\dag)=|S_0\rangle \langle S_1| (|S_1\rangle\langle S_0|)$ is the annihilation  (creation) operator for a molecular excitation, where $S_0$ and $S_1$ are the ground and electronic excited states, respectively. $\Omega(\bR)$ is the transition  energy between the $|S_0\rangle$ and $S_1$ states along the reaction coordinate $\bR$. $\bar{g}(\bR)=\mathbf{\mu}(\bR)\cdot \mathbf{E}$ is the strength of the light-matter interaction with $\mathbf{\mu}(\bR)$ being the transition dipole moment of the molecule between $S_0$ and $S_1$ states. $\mathbf{E}=E\mathbf{u}\sqrt{\omega_c/2\epsilon V_c}$ is the electric field, where $V_c$ is the effective mode volume of the cavity photon, $E$ is the amplitude of the photon electric field at the molecular position, $\mathbf{u}$ is the unit vector indicating the direction of the electric field of the cavity photon and $\epsilon$ is the dielectric constant.

\begin{figure}
 \centering
  \includegraphics[width=0.5\textwidth]{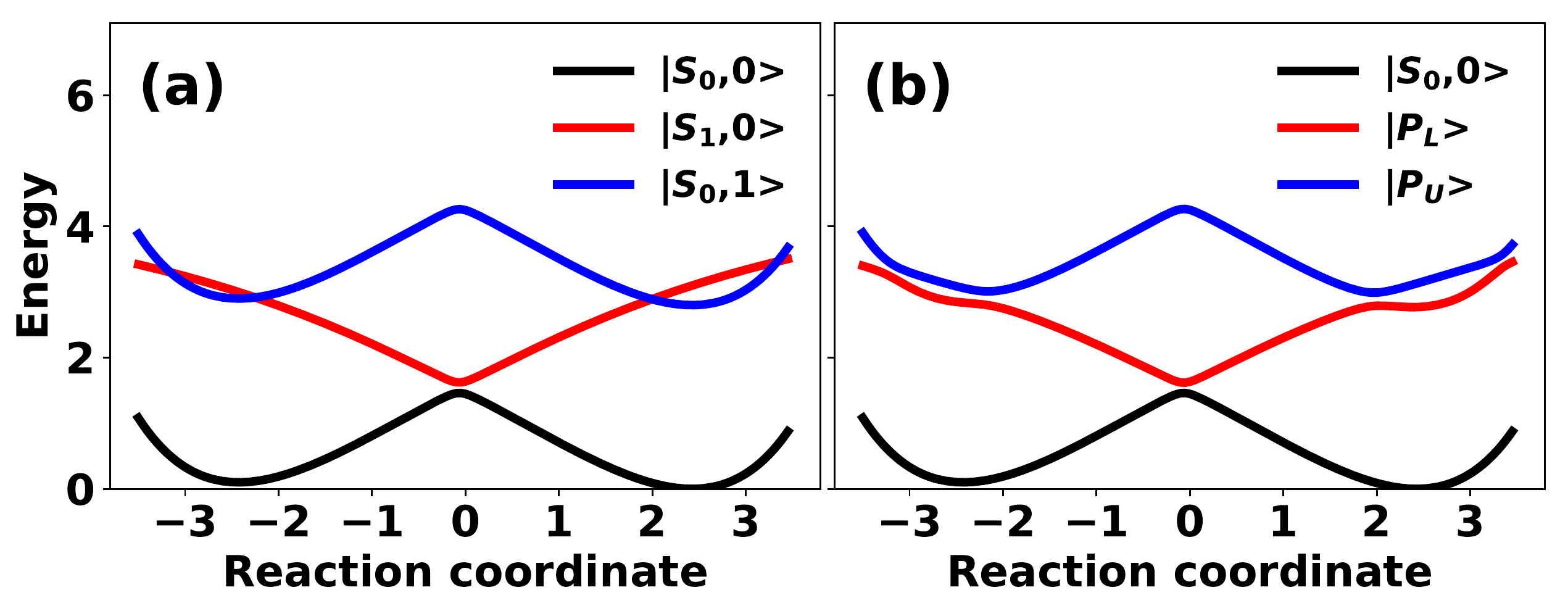}
  \caption{\label{fig1}Schematic diagram of Rabi splitting in PESs of molecular systems (mimic the PESs show in Fig.~2b of Ref~\onlinecite{Galego2016}). (a) Ground states (black), electronic (red) and photonically (blue) excited states before couplings. (b) The light-matter interactions lead to the hybridization between the $|S_0,1\rangle$ and $|S_1,0\rangle$, resulting in appearance of lower ($|P_L\rangle$) and upper ($|P_U\rangle$) polaritonic states.}
\end{figure}

The expression \emph{strong coupling} means that the coupling between light and matter is strong compared to the damping of each degree of freedom. The coupling term ensures that the number of excitations $\hat{N}=a^\dag a + \sigma^\dag\sigma$ is preserved in the basis set of $|S_{0/1},N_c\rangle$~\cite{C8SC01043A}, where $|N_c\rangle$ is the $N_c$ photon Fock state, $a|N\rangle=\sqrt{N_c}|N_c-1\rangle$. Diagonalization of the Hamiltonian in Eq.~\ref{eq:hjc} results in sets of polaritonic states: 
\begin{align}\label{eqplu}
|P_L\rangle =&  \sin\theta|S_0,N_c+1\rangle + \cos\theta|S_1,N_c\rangle,\nonumber\\
|P_U\rangle =&  \cos\theta|S_0,N_c+1\rangle - \sin\theta|S_1,N_c\rangle,
\end{align}
where $|S_0,N_c\rangle (|S_1,N_c\rangle)$ denotes a many-body state where the molecule is in the ground (excited) state and the cavity has $N_c$ photons. The corresponding energy along the reaction coordinates is shown by the blue (red) line in Fig.~\ref{fig1}(a) revealing that the photonic excited state (blue) is similar to the ground state (black) with energy shifted by the $\omega_c$. The light-matter mixing angle $\theta$ in Eq.~\ref{eqplu} satisfies $\tan(2\theta)=\frac{2g\sqrt{N_c}}{\Delta}$. The corresponding eigenvalues are
\begin{equation}\label{eqepm}
E_\pm(\bR) = \frac{\Omega(\bR)+\omega_c}{2}\pm \sqrt{\frac{\Delta^2(\bR)}{4}+\bar{g}^2(\bR)N_c},
\end{equation}
where $\Delta(\bR)=\Omega(\bR)-\omega_c$ is the molecular-photon detuning. The resulting polaritonic energies are illustrated by the red and blue lines in Fig.~\ref{fig1}(b). Thus the energy difference between the two polaritonic states is the Rabi splitting
\begin{equation}\label{rabi1}
\Omega_r(\bR)=2\sqrt{\Delta^2(\bR)/4+\bar{g}^2(\bR)N_c}.
\end{equation}
If $\Delta(\bR)$ is zero, Rabi splitting becomes $2 \bar{g}(\bR) \sqrt{N_c}$. Eq.~\ref{rabi1} and the comparison between Fig.~\ref{fig1}(a) and (b) indicate that when the detuning is large (compared to the coupling strength, i.e., $|\Delta(\bR)|\gg |\bar{g}(\bR)|$), the Rabi splitting approaches the energy difference between the two bare states. In this situation, the light-matter interaction has very limited effect on the electronic structure. In contrast, when $|\Delta(\bR)|\ll |\bar{g}(\bR)|$, $\tan(2\theta)\rightarrow \pm\infty$, the mixing of light and matter states in $P_{L/U}$ becomes significant. 

The JC model can be further generalized to include $M$ identical two-level emitters that interact strongly with a lossless cavity mode, which is known as the Tavis-Cummings (TC) model~\cite{PhysRev170379,PhysRev188692}. Similar to the JC model, the interaction only occurs within the subspace that the total number of excitations $\hat{N}=a^\dag a+\sum^M_i\sigma^\dag_i\sigma_i$ is perserved. Consequently, the hybridized states can only form from the basis states that have the same excitations $N_{exc}$. For the $N_{exc}=1$ excitation, there are $M+1$ basis states: photonically excited state $|G,1\rangle$ and single electronic excited state $|S^{(k)}_1,0\rangle$, where $|G\rangle$ represents the case where all molecules are in the ground state and $S^{(k)}_1$ indicates that only the $k^{th}$ molecule is in the excited state. Thus, in the $N_{exc}=1$ subspace,  the eigenstates of the TC model consist of lower and upper polaritonic states, formed by the hybridization between the electronic excited states $\frac{1}{\sqrt{M}}\sum_k |S^{(k)}_1,0\rangle$ and the photonically excited state $|G,1\rangle$, and $M-1$ dark states~\cite{C8SC01043A}. The energy gap between the two polaritonic states is similar to that of the JC model,
\begin{equation}\label{rabi2}
\Omega_r(\bR)=2\sqrt{\Delta^2(\bR)/4+g^2(\bR)N_cM}.
\end{equation}
The above equation indicates that the Rabi splitting is affected by the concentration of the molecules. In fact, concentration is usually a tunable parameter to observe strong Rabi splitting in experiments~\cite{nat2016strong}.

The TC and JC models have been extensively used to study strong coupling cases. However, these simplified models have limited applications in the prediction of complex dynamics in condensed systems with many degrees of freedom. In order to describe the effect of strong light-matter interaction across many chemical and physical processes in realistic systems, an atomistic model, is needed as suggested in the next section.  

\subsection{Electronic structure in the presence of strong light-matter interaction}
\label{sec:eslight}
When the coupling between a molecular system and a confined light mode is strong enough to overcome the dissipative processes, the light-matter interaction enters into the strong-coupling regime. As a result, the molecular system and photon are hybridized and new quantum states, polaritons, appear.
In the strong-coupling regime, one practical framework for solving the light-matter Hamiltonian is the RWA where the fast oscillating terms are neglected~\cite{qoptics} because 1) the light-matter coupling strength is still weak compared to the excitation energies, 2) The timescale of the fast oscillating terms, $\approx \frac{\hbar}{2\omega_c}$, is less than 1~fs, which is much shorter than the timescale we are simulating, 3) The $\bar{g}_i(\bR)\ll \omega_c$ is satisfied in our simulations.
The Hamiltonian of the molecule strongly interacting with the cavity can then be expressed as
\begin{align}\label{htot}
\hat{H}_{tot}&= \hat{H}_e+\hat{H}_c+\hat{H}_{int} \nonumber\\
=&\sum^N_{i=1} \Omega_i(\bR) c^\dag_i c_i + \omega_c a^\dag a
+ \sum^N_{i=1}\bar{g}_i(\bR)(c^\dag_i a + c_i a^\dag),
\end{align}
where $\hat{H}_e$, $\hat{H}_c$ are the electronic Hamiltonians of the molecule and cavity photon, respectively. $\hat{H}_{int}$ describes the interaction between an electron and cavity photon. $N$ is the number of excited states of interest. $c_i (c^\dag_i)$ is the annihilation (creation) operator of the $i^{th}$ electronic state and $\Omega_i(\bR)$ is the corresponding excitation energy of the molecule. The ground state energy is defined as $E_g(\bR)$, thus the total energy of the $i^{th}$ excited state is $E_g(\bR)+\Omega_i(\bR)$. The ground state energy and excitation energies of the molecular system can be calculated using many standard quantum-mechanical methods, including first-principles and semi-empirical techniques. It is important to note that, in present work, we set a tunable parameter $g$ for the electric field that controls the light-matter coupling strength, i.e., $\bar{g}_i(\bR)=g \mu_i(\bR)\cdot \bu$, where $\mu_i(\bR)$ is the transition dipole of the $i^{th}$ electronic excited state.

After excitation energies, $\Omega_i(\bR)$, and coupling strengths, $\bar{g}_i(\bR)$, are obtained, the energies of the polaritonic states can be readily calculated by diagonalizing the Hamiltonian, Eq.~\ref{htot}, in the basis set of light-matter states, $\{|S_i,0\rangle,|S_0,1\rangle\}$,
\begin{equation}\label{cavityH}
H=
\begin{bmatrix}
\omega_c &  g_1  & \cdots &  g_N  \\
 g_1   & \Omega_1(\bR) & \cdots & 0          \\
 \vdots     & \vdots     & \ddots & \vdots     \\
 g_N   & 0          & \cdots & \Omega_N(\bR) \\
\end{bmatrix}.
\end{equation}
Alternatively, one may use the Floquet theory~\cite{RMP89011004} to describe a molecule-light coupled system in a semiclassical way, which has been employed in several quantum molecular dynamics~\cite{Kim_2015,Hal_sz_2011,jp206860p,KOROLKOV20041}. Diagonalization of Eq.~\ref{cavityH} results in the polaritonic states with total energies $E_K$ (shifting the eigenvalue by $E_g(\bR)$) and wave function (WF) 
\begin{equation}\label{polaritonwf}
\Psi_K= \sum^{N}_{i=0}\beta^K_i |\psi_i\rangle,
\end{equation}
where $\beta_i$ is the $i^{th}$ eigenvector of $H$. $|\psi_0\rangle=|S_0,1\rangle$ is the photonically excited state and $|\psi_i\rangle=|S_{i},0\rangle$ for $i>1$ is the molecular excited state. The index $K$ runs over the $N+1$ eigenstates of the system. As shown by Eq.~\ref{polaritonwf}, the polaritons are formed by the superposition of the molecular states with the single-photon state. If there are $M$ molecules, at least $M$ degenerate $S_1$ states are present. The superposition between $M$ degenerate $S_1$ states and the single-photon state leads to the formation of polaritonic states delocalized over the $M$ molecules, which may in return affect many chemical and physical properties, including charge transfer, energy transfer, and collective photochemical reactions~\cite{Coupling2017,Orgiu2015,PRL119136001}.

These processes may be captured via MQC family of approaches, where typically classical trajectories of all atoms in the system are computed by numerically integrating Newton's equations of motion associated with the PESs. Even though the force constants of the polaritonic state may be quite different from those of bare molecules, the derivatives of the polaritonic PES $\frac{\partial E_i(\bR)}{\partial \bR}$ can be related to the derivatives of the bare PESs through a unitary transformation according to Eq.~\ref{polaritonwf}. For atom $a$ of the molecule in the polaritonic state $K$, the force $F^K_a= -\nabla_a E_K$ is
\begin{align}\label{eqforce}
F^K_a= &-\langle\Psi_K|\nabla_a \hat{H}|\Psi_K\rangle 
= -\left(\sum^{N}_{i=0} \beta^{*K}_i \langle \psi_i| \right)\nabla_a \hat{H} \left(\sum^{N}_{j=0} \beta^K_j |\psi_j\rangle   \right) \nonumber\\
= & -\sum^N_{i=1} (\beta^K_{i})^2 \nabla_a\langle S_i| \hat{H}|S_i\rangle
-(\beta^K_0)^2 \nabla_a \langle S_0|\hat{H}|S_0\rangle \nonumber\\ &
-\sum^N_{i=1} \beta^{*K}_0\beta^K_{i} \nabla_a \langle S_0|\hat{H}| S_i\rangle
-\sum^N_{i=1} \beta^K_{i}\beta^{*K}_o \nabla_a \langle S_i|\hat{H}| S_0\rangle .
\end{align}
According to the above equation, derivatives of the bare PES and the respective bare transition dipole, $\mu_i(\bR)$, are required in the calculation of forces on the polaritonic PES. Hence, excited-state properties of the bare molecule must be calculated. 

\subsection{Gradients and non-adiabatic couplings in TDSCF framework}
\label{sec:gradients}
Compared to the ground state calculations, the calculations of electronic excited state properties, including energies, gradients, and transition dipoles, are more complex due to the presence of many-body interactions. Many methods have been developed to treat these interactions at different levels of accuracy and efficiency. The time-dependent self-consistent field (TDSCF) framework is one of the more computationally efficient and popular methods~\cite{jcp4905828,C3CP51514A,jcp1508368}. TDSCF is a general method for excited state properties and can be implemented in the form of time-dependent density functional theory (TDDFT)~\cite{C3CP51514A,jcp1508368} or time-dependent Hartree-Fock (TDHF) or Configuration Interaction Singles (CIS) methods~\cite{jcp4905828}. In our work, we use the semiempirical formulation of the TDSCF framework~\cite{jcp4905828} as realized in the NEXMD software package~\cite{ar400263p}. Within the TDSCF framework, the excited energies $\Omega_i$ and the respective transition density matrix between the ground state and $i^{th}$ excited state $\xi_i$ can be calculated on-the-fly using Krylov space methods~\cite{DAVIDSON197587}, where $\xi_{i,\mu\nu}=\langle\psi_0|c^\dag_\mu c_\nu| \psi_i\rangle$ is in the atomic basis spanned by indices $\mu$ and $\nu$~\cite{qua1528}. The gradients of excited energies can be calculated analytically as well~\cite{jp109522g,C3CP51514A}.The method has been demonstrated to be efficient for calculating the excited state properties, including energies and analytical gradients, of realistic molecular systems~\cite{jp109522g}. 

After the transition density matrix is calculated, the transition dipole $\mu_i(\bR)=\text{Tr}[\bar{\mu}(\bR)\xi_i]$, is the respective expectation value for the dipole operator represented as a matrix $\bar{\mu}(\bR)=\langle\mu|\bR|\nu\rangle$. Thus the derivative of the transition dipole is
\begin{equation}
\frac{\partial\mu_i}{\partial \bR_a}= \sum_{\mu\nu}\left[\frac{\partial  \langle\mu | \bR|\nu\rangle}{\partial \bR_a}
\xi_{i,\nu\mu}+\langle \mu|\bR|\nu\rangle \frac{\partial \xi_{i,\nu\mu}}{\partial\bR_a}\right],
\end{equation}
where $|\mu\rangle, |\nu\rangle$ are atomic orbitals. 
The calculation of the first term in the above equation is trivial. However, the second term requires the derivative of the respective transition density matrix. Following the derivation in Ref.\citenum{qua1528}, the derivative of transition density can be obtained analytically by applying $\partial/\partial \bR$ to the TDSCF eigen-equation $\mathcal{L}\xi_i=\Omega_i\xi_i$. The tedious derivation can be found in Appendix.~I. 

The non Born-Oppenheimer effects and non-adiabatic transitions between different electronic states are determined by so-called derivative Non-Adiabatic Couplings (NACs). The non-adiabatic coupling vector, termed as NACR, is defined as
\begin{equation}
\bd_{KL}=\langle \Psi_K| \nabla_\bR\Psi_L\rangle.
\end{equation}
Notably, NAC is also dressed by the photonic state after the formation of polaritonic states. When strong light-matter interaction is present, the excited states and corresponding WFs are altered as given by Eq.~\ref{polaritonwf}. Consequently, the NAC is modified by the strong light-matter interaction according to the derivation shown in Appendix.~II. The NACR is then written in Hellmann-Feynman like form as
\begin{equation}\label{eqnac1}
\bd_{KL}=\langle \Psi_K| \nabla_\bR\Psi_L\rangle
=\frac{\langle \Psi_K(t)|\hat{H}^\bR|\Psi_L(t)\rangle}{E_L(t)-E_K(t)}.
\end{equation}
With the polaritonic WF defined by Eq.~\ref{polaritonwf}, the matrix element of the derivative of Hamiltonian, $\hat{H}^\bR$, in polaritonic states can be written as
\begin{align}\label{eqnacr}
\langle \Psi_K|\hat{H}^\bR|\Psi_L\rangle = & \left( \sum^N_{i=0} \beta^{*,K}_i \langle \psi_i| \right) 
\hat{H}^\bR\left( \sum^N_{j=0} \beta^{L}_j | \psi_j\rangle\right) 
\nonumber\\=&
\sum^N_{i,j=0}\beta^{*,K}_i\beta^L_j \text{Tr}[\rho_{ij}h^\bR],
\end{align}
where $(\rho_{ij})_{\mu\nu}=\langle \psi_i|c^\dag_\mu c_\nu|\psi_j\rangle$ is the single electron transition density matrix of the bare molecule between electronic states and $\hat{H}^\bR=\sum_{\mu\nu}h^\bR c^\dag_\mu c_\nu$. Substituting the above equation into Eq.~\ref{eqnac1} returns the NACR for strong light-matter interaction. Notably, for dynamical trajectories, the time-dependent non-adiabatic coupling (NACT) analog, $\bR\cdot\bd_{KL}=\frac{\partial\bR}{\partial t}\cdot\langle \Psi_K| \nabla_\bR\Psi_L\rangle=\langle \Psi_K|\nabla_t\Psi_L\rangle$, can be calculated in a similar way (Eqs.~A14-15 in appendix), i.e., 
\begin{equation}\label{eqnact}
\dot{\bR}\cdot\bd_{KL}=\langle \Psi_K| \nabla_t\Psi_L\rangle
=\frac{\langle \Psi_K(t)|\hat{H}^t|\Psi_L(t)\rangle}{E_L(t)-E_K(t)}.
\end{equation}

In summary, the above formalism outlines the calculation of gradients and NAC between polaritonic states within the TDSCF quantum-chemical approaches for excited states, reducing the problem to calculation of derivatives of the Hamiltonian matrix elements for the bare molecule, within the framework of analytic derivative techniques~\cite{jp109522g}. Adaptation of this approach to analytic derivative methods developed for TDDFT is straightforward~\cite{C3CP51514A}. Moreover, it should be noted that the NAC between the polaritonic states can be easily calculated from the density matrix and derivative of the Fock matrix of the bare molecule by a unitary transformation. Alternatively, the NACT value $\langle \Psi_K|\frac{\partial}{\partial_t}|\Psi_L\rangle$ is frequently approximated numerically as $\langle\Psi_K(t)|\Psi_L(t+\Delta t)\rangle/\Delta t$ in accordance with Hammes-Schiffer and Tully's proposal~\cite{jcp467455,jctc7b00388}. Previous computational studies assumed that the bare adiabatic states vary slowly, and thus approximated the overlap $\langle\Psi_K(t)|\Psi_L(t+\Delta t)\rangle$ as $\sum^N_j \beta^{*K}_j(t)\beta^L_j(t+\Delta t)$~\cite{jctc7b00388}. In this case, the overlaps between the bare states are assumed to be $\langle \psi_i(t)|\psi_j(t+\Delta t)\rangle\approx\delta_{ij}$. It should be noted that  this is true only when one electronic excited state is considered, since $\langle \psi_0(t)|\psi_1(t+\Delta t)\rangle=\langle S_0(t),1|S_1(t+\Delta t),0\rangle=0$ due to the different photon state~\cite{jpclett9b01599}. However, if multiple electronic excited-states are included, the overlaps, $\langle \psi_i(t)|\psi_j(t+\Delta t)\rangle=\langle S_i(t),0|S_j(t+\Delta t),0\rangle$ ($i,j>0$), are the NACTs of the bare molecule and should be nonzero for $i\neq j$ in order to ensure non-adiabatic transition between the bare electronic states.

\subsection{Non-adiabatic molecular dynamics with strong light-matter interaction}
\label{sec:namd}
Excited energies, gradients, and NACs are the basic ingredients for non-adiabatic  molecular dynamics (NAMD). In MQC methods, NAMD is implemented as an approximate solution of the Schr\"odinger equation (TDSE) for both electrons and nuclei. Here, the electronic and nuclear degrees of freedom are separated. Fast electronic dynamics are subject to the TDSE for electrons while slow nuclear motions are treated classically. As a result, electronic properties (including potential energies, gradients, and non-adiabatic couplings) are computed at each time step of a trajectory. The MQC treatment of NAMD significantly reduces the computational cost, and thus can be implemented to compute electronic quantities on-the-fly~\cite{cr7b00577}. Across various NAMD approaches, mean-field Ehrenfest dynamics~\cite{jcp437910,jcp2008258} and FSSH algorithm~\cite{jcp459170} are widely used. 

\subsubsection{Mean-field Ehrenfest dynamics}
\label{secmfed}
We start with the TDSE for both electrons and nuclei,
\begin{equation}\label{tdse}
i\hbar\frac{\partial }{\partial t}\Theta(\br,\bR,t)=\hat{H}\Theta(\br,\bR,t),
\end{equation}
where $\Theta$ is the total WF and $\hbar$ is the reduced Planck constant. The full Hamiltonian in this equation  is $\hat{H}=\hat{T}_n+\hat{H}_e$, where $\hat{T}_n$ is the kinetic energy operator for the nuclei and $\hat{H}_e$ is the Hamiltonian for the electrons. 

In Ehrenfest dynamics, the molecular WF is factorized as
\begin{equation}\label{factorization}
\Theta(\br,\bR,t)=\chi(\bR,t)\Phi(\br,\bR,t)\exp\left(\frac{i}{\hbar}\int^t_{t_0}dt' E_e(t')\right),
\end{equation}
where the phase factor is determined by the total electronic energy $E_e=\langle \Theta|\hat{H}_e|\Theta\rangle$. $\chi$ is the WF of the nuclei. For bare molecules, $\Phi(\br,\bR,t)$ is the electronic WF depending on the nuclear coordinates. In the presence of light-matter interaction, $\Phi(\br,\bR,t)$ is the polaritonic WF calculated via Eq.~\ref{polaritonwf}. Applying the factorization in Eq.~\ref{tdse}, the TDSE can be projected in the polaritonic and nuclear spaces leading to two coupled equations of motion (EOMs) for $\chi$ and $\Psi$. After the decomposition, the polaritonic WF $\Phi(\bR,t)$ is determined by the following TDSE,
\begin{equation}\label{etdse}
i\hbar\frac{\partial}{\partial t}\Phi(\br,\bR,t)=\hat{H}_e\Phi(\br,\bR,t).
\end{equation}
where the electronic WFs depend on the classical nuclear coordinates. One advantage of the Ehrenfest dynamics is that the WF $\Phi$ can be directly propagated using Eq.~\ref{etdse} without choosing any basis functions. If expansion of the WF in terms of basis functions is desired, $\Phi$ can be expanded as
\begin{equation}\label{expansion}
\Phi(\bR,t)=\sum_K c_K(t)\Psi_K(\bR(t)),
\end{equation}
where $c_i(t)$ are the time-dependent expansion coefficients. As a result, the quantum EOM (Eq.~\ref{etdse}) can be reduced to
\begin{equation}\label{ceom}
i\hbar\dot{c}_K=\sum_L \left[H_{KL}-i \dot{\bR}\cdot \bd_{KL}\right]c_L.
\end{equation}
Here $H_{KL}=\langle \Psi_K|\hat{H}_e|\Psi_L\rangle$ are matrix elements of the Hamiltonian $\hat{H}$ in the basis functions. $H$ is diagonal if adiabatic states are employed in Eq.~\ref{expansion}. 

The EOM for $\chi$ in the classical limit is equivalent to Newton's equation for nuclei on the average potential of electrons~\cite{A801824C}.
\begin{equation}\label{neom}
M_a\ddot{\bR}_a= -\nabla_a \langle \Phi(\bR)|\hat{H}_e|\Phi(\bR)\rangle 
=-\sum_{KL} c^*_K c_L \langle \Psi_K|\nabla_a\hat{H}_e|\Psi_L\rangle,
\end{equation}
where $M_a$ and $\ddot{\bR}_a$ are the mass and acceleration, respectively, of the $a^{th}$ nucleus. The classical equation of motion (EOM) in Eq.~\ref{neom} can be integrated with standard methods, such as velocity Verlet algorithm~\cite{verlet1998}. The quantum EOM in Eq.~\ref{etdse} is solved along the classical trajectories. From Eqs.~\ref{ceom} and ~\ref{neom}, it can be derived that the quantum and classical EOMs in the presence of light-matter interactions are the same as the counterparts for bare molecules except the energies, WFs, forces, and non-adiabatic couplings are dressed by the light-matter interaction. 

As prescribed by Eq.~\ref{neom}, the evolution of classical nuclei is subject to an effective potential corresponding to an average over quantum states. This is also the reason why the Ehrenfest dynamics is a type of mean-field method. This mean-field treatment results in forces averaged from more than one PES. Consequently, the Ehrenfest dynamics may not correctly represent different physical situations when the dynamics leaves regions of strong NAC. For example, the Ehrenfest trajectories of low-probability paths are quite similar to the dominating trajectories, leading to poor representation of low-probability events. Moreover, Ehrenfest dynamics does not satisfy the principle of detailed balance, meaning that the forward and backward processes violate the microscopic reversibility~\cite{jcp4757762}. 

\subsubsection{Fewest switches surface hopping}
Within FSSH, the NAMD is modeled via propagation of a swarm of classical trajectories. Each trajectory evolves on a single adiabatic PES with the respective gradients. The non-adiabatic effects come from the trajectory hopping between the PESs in a stochastic way. Once the forces and NACs are known, the NAMD can be readily performed by integrating the Newton or Langevin equations. The hopping between different PESs is determined by Tully's FSSH algorithm~\cite{jcp459170}. In particular, the EOM of $c_K(t)$ can be straightforwardly derived from the Schr\"odinger equation in the adiabatic polaritonic eigenstates $\Psi_i$, 
\begin{equation}\label{fssheom}
i\hbar\dot{c}_K(t)=c_K(t)E_K(t)-i\sum_L c_L(t)\dot{\bR}\cdot \bd_{KL},
\end{equation}
where $\dot{\bR}\cdot \bd_{KL}$ is the NACT calculated by Eq.~\ref{eqnac1}. With the propagation of the EOM for $c_k(t)$, the matrix elements of the time-dependent density matrix in the adiabatic eigenstates is given by $\sigma_{KL}=c^*_{K}(t)c_L(t)$. The diagonal and off-diagonal elements of $\sigma$ represent the occupation probabilities of adiabatic states and coherence between adiabatic states, respectively. It should be noted that the spontaneous emission of the quantum system is not included in Eqs.~\ref{ceom} and ~\ref{fssheom}. Because the spontaneous emission occurs on a much longer time scale (ns) compared the non-radiative decay process (ps), it is safe to ignore the spontaneous emission in the time-scale (ps) we are interested in to model photochemical reactions.

Nuclear motion along the polaritonic PES is subject to a constant temperature Langevin dynamics or Newtonian dynamics, 
\begin{equation}\label{langevin}
M_a\ddot{\bR}_a(t)=-\nabla_a E_K(\bR(t))-\gamma M_a\dot{\bR}_a(t)+\bA(t),
\end{equation}
where $\dot{\bR}_a$ is the velocity of the $a^{th}$ nucleus. The first term on the right-hand side of the above equation is the force on the polaritonic PES calculated from Eq.~\ref{eqforce}. $\bA(t)$ is the stochastic force which depends on the bath temperature and $\gamma$ is the corresponding friction coefficient. The Langevin equation reduces to Newtonian motion when the stochastic force and damping coefficient are both zero. Again, as argued in Sec.~\ref{secmfed}, the quantum and classical EOMs in Eqs~\ref{fssheom} and ~\ref{langevin} are the same as the counterparts without the light-matter interaction except the forces, energies, and non-adiabatic couplings are dressed by the light-matter interaction. As with Ehrenfest dynamics, the classical EOM , Eq.~\ref{langevin}, is solved by velocity Verlet integration~\cite{verlet1998}. The energies, and gradients in Eq.~\ref{langevin} are calculated on-the-fly at every trajectory point $\bR(t)$. 

In FSSH, the probability of transition between adiabatic states $K$ and $L$ during a time interval depends on the time-dependent density matrix and NACT. Since electronic motion is faster than nuclei, the quantum time-step $\delta t$ used to propagate the quantum coefficients is smaller than the classical time-step $\Delta t$ for Langevin or Newtonian dynamics. At every classical time step, the switching probability $P_{KL}$ is evaluated as the summation over all $N_q=\Delta t/\delta t$ quantum steps,
\begin{equation}\label{gkl}
P_{KL}=\frac{\sum^{N_q}_jb_{LK}(j)\delta t}{\sigma_{KK}}.
\end{equation}
where $b_{LK}=-2\text{Re}(\sigma^*_{KL} \dot\bR\cdot \bd_{KL})$. The hops between adiabatic states are accepted or rejected stochastically according to the calculated hopping probability and uniformly generated random number~\cite{ar400263p}. Following the hop, nuclei evolve on the PES of the new state. Energy is conserved through rescaling nuclear velocities along the direction of NACR. If the nuclear kinetic energy is not sufficient to hop to the higher energy state, the hop is rejected. Finally, it should be noted that all improvements of the original FSSH model developed for isolated molecules, such as decoherence corrections~\cite{jcp4809568}, trivial crossings~\cite{NELSON2013208,jz500025c}, are trivially  extendable to polaritonic FSSH presented here.

\subsection{Numerical implementation and applications}
\label{numerical}
The FSSH scheme of NAMD for strong light-matter interaction has been implemented in the NEXMD code~\cite{ar400263p} used for all simulations described below. NEXMD has been broadly used for modeling of photoinduced dynamics in a variety of molecular systems~\cite{Nelson2018,jctc8b00103,jpca5b02092,jp109522g}.

\begin{figure}
\centering
  \includegraphics[width=0.36\textwidth]{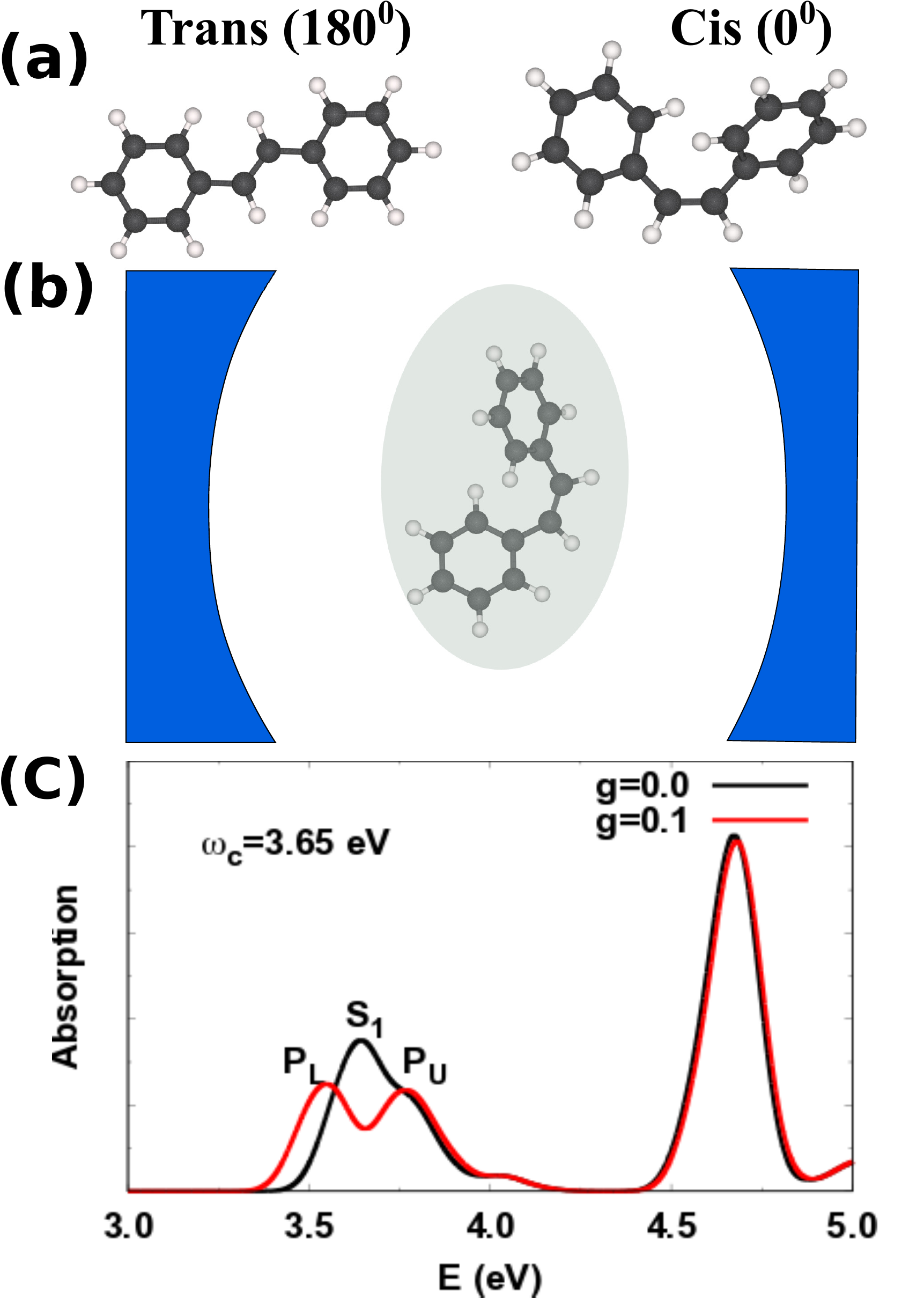}
  \caption{\label{fig2} (a): Geometries and dihedral angles of \textit{trans}- and \textit{cis}-isomers of the stilbene. (b): Schematic diagram of this molecule in cavity. The dash curves demonstrate the photonic mode supported by the cavity. Absorption spectrum of \textit{cis} isomer with (red) and without (black) light-matter interaction. The photon energy is chosen as 3.65~eV in resonance with the first absorption peak ($S_1$) of the bare \textit{cis} isomer. After the light-matter interaction is turned on, the first absorption peak is split into two peaks which correspond to the lower ($P_L$) and upper ($P_U$) polaritonic states.}
\end{figure}

As a case study for strong light-matter interaction (Fig.~\ref{fig2}(b)), we modeled the photoisomerizaiton of stilbene (Fig~\ref{fig2}(a)) to demonstrate the utility of the methodology. Photoisomerization is one of the most important photochemical reactions, which has found many applications in many fields, including energy storage and photoswitches~\cite{C1EE01861B,cr500249p}. However, it has to be suppressed in certain cases~\cite{C3CP54048K}. In order to control the photoisomerization, strong coupling with quantized light modes can be explored~\cite{Galego2016,Fregoni2018}. Fig.~\ref{fig2}(a) sketches the structure of \textit{trans} and \textit{cis} stilbene isomers. The interconversion between these two isomers is a commonly studied photoisomerization reaction~\cite{cr00003a007}. The effect of strong light-matter interaction on the isomerization reaction of the stilbene has been studied in previous work by employing a quantum non-adiabatic dynamics method~\cite{Galego2016}. But the computational complexity of quantum non-adiabatic dynamics limits the simulation to a simple one-dimensional model.

The geometries of the \textit{trans} and \textit{cis} isomers were optimized at the AM1 semiempirical level~\cite{ja00299a024} coupled with the CIS approach for excited states using our NEXMD package~\cite{ar400263p}. The optimized structures were used as the starting points for two separate ground state Born-Oppenheimer Molecular Dynamics (BOMD) trajectories (one trajectory each for \textit{trans} and \textit{cis} species) performed using Langevin dynamics at room teperature (300~K) with a friction coefficient of 20~ps$^{-1}$. After initial heating and equilibration (300~K), 1~ns of constant temperature ground state BOMD simulations were carried out with a time-step of 0.5~fs at the Hartree-Fock level. Each equilibrated trajectory was used to collect a set of 500 initial coordinates and momenta for all subsequent NAMD simulations for the corresponding species. For each of the configurations, single point calculations were performed using the CIS technique~\cite{Mukamel781}  to determine vertical excitation energies and oscillator strengths for the 5 lowest excited states. Oscillator strengths were Gaussian-broadened at the excitation energies with an empirical standard deviation of 0.05~eV to obtain the absorption spectrum of each configuration, i.e.,  $A(\omega)=\sum_K \frac{f_K}{\sqrt{2\pi\Gamma}}\exp\left[-\frac{(E_K-\omega)^2}{\Gamma^2}\right]$, where $\Gamma$ is the Gaussian-broadening and $f_K$ is the oscillator strength of the $K^{th}$ state. Since the polaritonic states are the superposition of the bare molecular states, the oscillator strengths of the polaritonic states can be calculated from the bare oscillators with a unitary transformation constructed by the eigenvector of Eq.~\ref{cavityH}. From that, the absorption spectrum is calculated by averaging over all absorption spectra of each configuration. The calculated absorption spectrum for the \textit{cis}-isomer of stilbene with and without light-matter interaction is shown in Fig.~\ref{fig2}(c). As expected, when the molecule is resonantly interacting with the photon in a cavity tuned to the excitation energy of $S_1$ state at 3.65 eV, this state splits into  $P_L$ and $P_U$ polaritonic components (Fig.~\ref{fig2}(c)) as discussed in detail below. 

For the NAMD on the PES of excited states, each simulation is carried out for 1~ps, including a total of 5 excited states with initial excitation to the $S_1$ (without light-matter interaction) or $P_U$ (with light-matter interaction) states. The classical nuclei were propagated using a 0.1~fs time step. The electronic quantum dynamics were solved with an intermediate time step of 0.025~fs~\cite{jp109522g}. The quantum time step is reduced by a factor of 10 near trivial crossings~\cite{jcp4732536}. 

\subsubsection{Photoisomerization of bare stilbene}

\begin{figure}
  \includegraphics[width=0.48\textwidth]{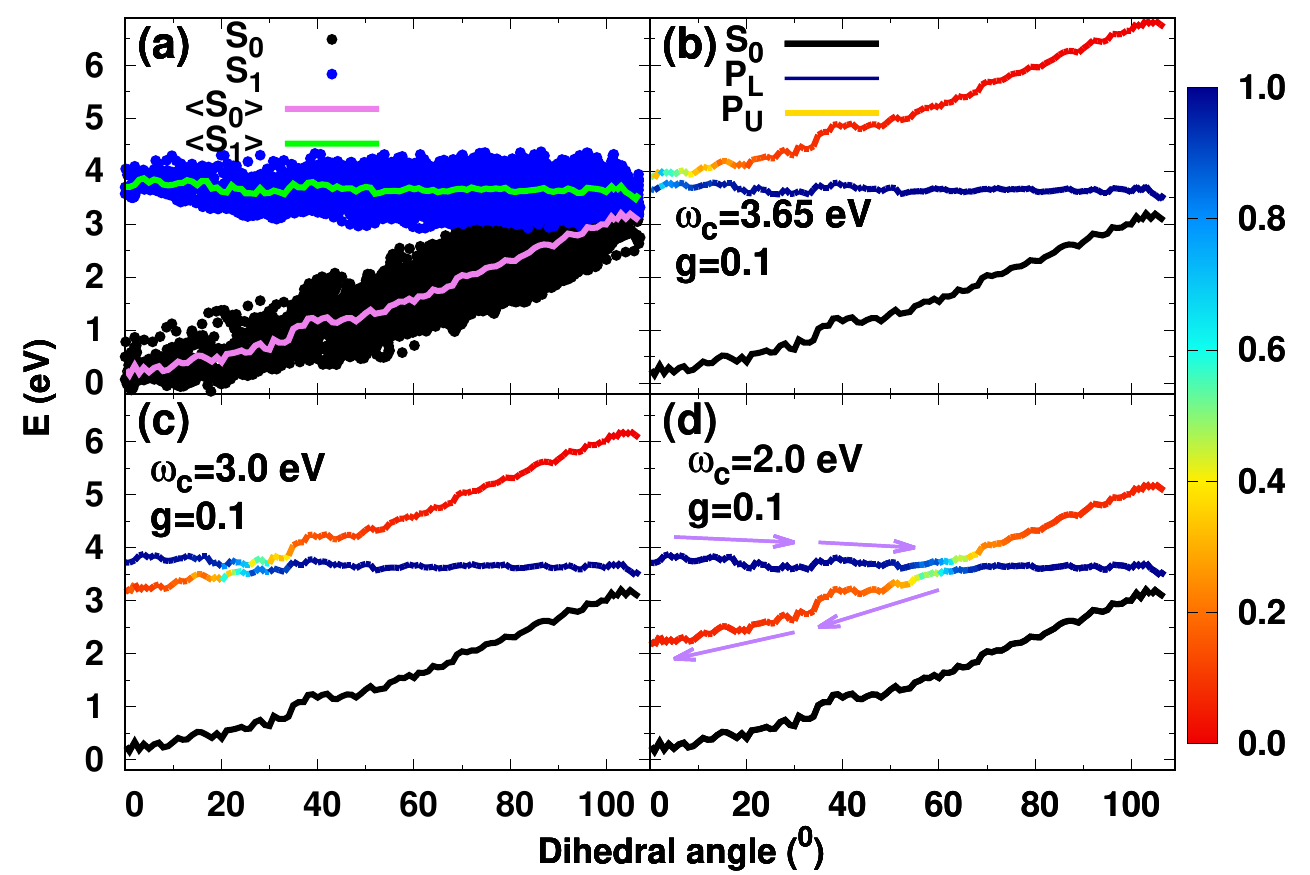}
  \caption{\label{figpes}PESs of \textit{cis} isomer without (a) and with (b-d) light-matter interaction. a) The black and blue dots are ground and excited state energies of different geometries. The red and green lines sketch the average ground and excited PESs along the reaction coordinate, respectively. b-d): ground (black) and excited-states (red-blue color scale) PESs of the molecule interacting with a photonic mode with energy being b) 3.65~eV, c) 3.0~eV,  and d) 2.0~eV, respectively, and $g=0.1$. The color scale represents the contribution of photonically excited state (red) and bare excited state (blue) to the hybridized excited-states.}
\end{figure}

The dihedral angle through the central $C=C$ bond ($\phi_{C=C}$) represents the reaction coordinate relevant for isomerization. The BOMD on the $S_1$ PES starting from the \textit{cis} configuration is able to scan the reaction coordinate from $0^0$ to the conical intersection (CI) (located near $\phi_{C=C}\simeq 90^0$). Then, the electronic structure calculations on the snapshots taken from the $S_1$ BOMD trajectory show the evolution of PESs along the reaction coordinate as shown in Fig.~\ref{figpes}(a). The same snapshots are also used to calculate the polaritonic PESs shown in Fig.~\ref{figpes}(c-d). The PES of this model structure shows that the ground state PES has a minimum at a dihedral angle of $0^0$ corresponding to the \textit{cis}-configuration. The blue dots in Fig.~\ref{figpes}(a) indicates that the PES of the first excited state fluctuates around 2.8$\sim$3.8 eV along the reaction coordinate. The green line sketches the average excited energy along the reaction coordinate. Thus, the PESs shown by Fig.~\ref{figpes} indicate that the \textit{cis-trans} transition can be readily induced by exciting the molecule to the first excited state. The photoinduced isomerization process involves passage through the CI between the excited state and ground state (S$_1\rightarrow S_0$). The adiabatic dynamics on $S_1$ with a swarm of trajectories initialized with \emph{cis} configurations also confirms that the isomerization can readily be driven on the S$_1$ surface. Numerical simulations reveal that the change in the potential energy manifests itself in rotation around the $\phi_{C=C}$ dihedral angle and that the CI can be approached. Even though the limitation of adiabatic dynamics forbids passage through the intersection, the simulations confirm that the following \textit{cis-trans} isomerization may appear after passing through the CI. The wavepacket that encounters the CI undergoes a nonadiabatic transition to the ground state and follow the ground state PES to reach the other equilibrium configuration. The terminology ``wavepacket" is used to represent an ensemble of trajectories that has a distribution of momenta, can branch between different PESs, reflect, decohere, etc. ``Wavepacket on the upper surface" indicates the subset of trajectories that undergo the respective process and its dynamics are collective. The BOMD simulations starting from \textit{trans} found that the photoisomerization can also be induced by exciting the \textit{trans} configurations to $S_1$ states. However, $S_1$ states of \textit{trans} configurations have lower energies, which results in longer isomerization timescales which is consistent with experimental measurements~\cite{cr00003a007}. Consequently, the photoisomerization is simulated by starting from the \textit{cis} configurations in the remainder of this work.

In the present example, the non-adiabatic transition to $S_0$ is not explicitly simulated due to the inability to describe crossings between a multi-reference excited state (CIS) and a single reference Hatree-Fock ground state~\cite{ar400263p,TM2006}. Moreover, here we would like to present a clear case when photochemistry changes due to modification of related PES due to strong light-matter interactions (as opposed to other possible reasons such as variation of NACs, etc.). Therefore, we apply a simple model based on energy gaps to qualitatively describe $S_1\rightarrow S_0$ transitions, following Ref.~\citenum{jpca8b09103,jpcl6b02037}. Because NAC scales inversely with energy gap, transitions are likely to occur at geometries near the CI. In addition, the transition between $S_1$ and $S_0$ states can occur at finite energy gaps before reaching the CI. We set a $\Delta E=1.0~$eV threshold on the energy gap to qualitatively describe the rate at which trajectories approach the CI and subsequently transition to $S_0$. The quantum yield (QY) is defined as the fraction of trajectories evolving on $S_1$ that encounter an energy gap smaller than 1.0~eV. It should be noted that this method overestimates the QY because a) The nonadibatic transition to the ground state is more likely to happen at gaps smaller than 1.0~eV. Setting a smaller threshold reduces the QY as shown by Fig.~\ref{fig-delta}. However, the $S_0/S_1$ conical intersection may not be accurately described by the NEXMD.  Thus, a relatively large threshold of 1.0~eV is used in this work to ensure the molecular dynamics are far from the conical intersection~\cite{jpca8b09103}. b) realistically, transitions to the ground state should not be guaranteed even when the dynamics approaches an energy gap smaller than $\Delta E$, but instead should occur with some probability. c) After the transition to $S_0$, there is a possibility that the dynamics can propagate back to the original excited configurations (\textit{cis}) that is neglected in this approach. 

\begin{figure}
  \includegraphics[width=0.42\textwidth]{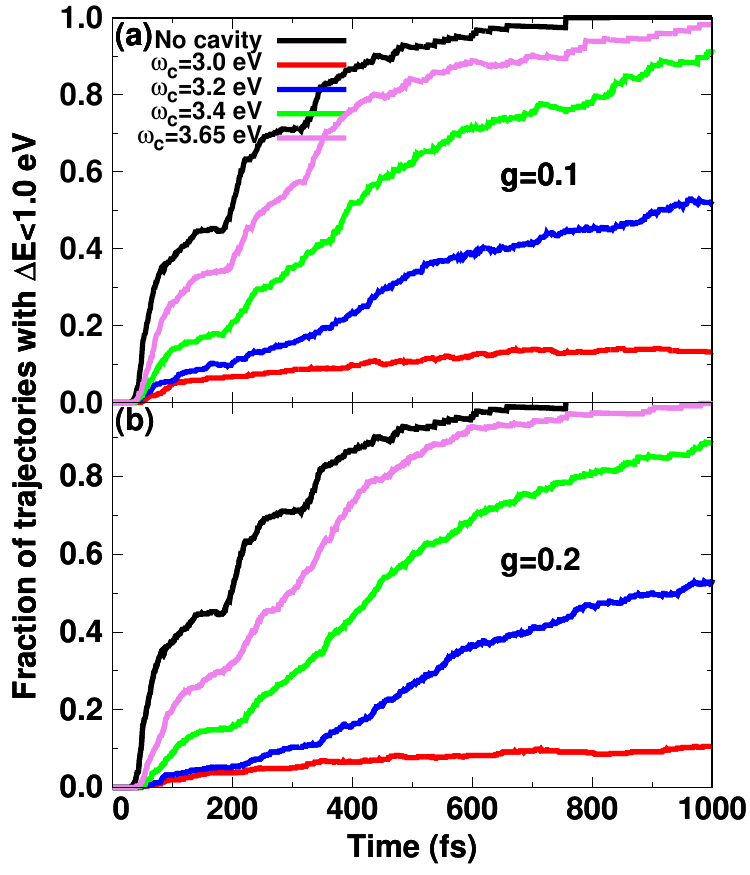}
  \caption{\label{figfrac}
Fraction of trajectories evolving on $S_1$ that encountered an energy gap $\Delta E<1.0$~eV for $g=0.1$ (a) and $g=0.2$ (b), respectively. The fraction of trajectories that can encounter an energy gap smaller than 1.0~eV decreases with lower photon energies.}
\end{figure}

In the absence of light-matter interaction, the PES of $S_1$ (blue dots and green line in Fig.~\ref{figpes}(a)) shows increasing energy fluctuations with increasing dihedral angle from 0 to 90$^0$ where the CI between $S_0$ and $S_1$ is located. Consequently, starting from the $S_1$ surface of \textit{cis} configuration, the photoisomerization from \textit{cis} to \textit{trans} can be induced by moving along the $S_1$ PES. The analysis on 500 trajectories found that the QY of photoisomerization (from \textit{cis} to \textit{trans}) is close to 100~\% after about 750~fs dynamics as shown in Fig.~\ref{figfrac}. Such simplistic definition of photoisomerization efficiency allows us to directly demonstrate its modifications due to polaritonic effects. The time evolution of the fraction of trajectories with $\Delta E<1$~eV shows a stepwise increase. Since there is no transition between $S_1$ and $S_0$ states, the dynamics may oscillate along the reaction coordinates ($0^0\rightarrow 90^0$). Fig.~\ref{fig-dihedral} plots the evolution of the mean dihedral angle, which clearly shows the oscillation. The oscillation period ($\sim150$~fs) corresponds to the time-interval between the steps shown by the black curve of Fig.~\ref{figfrac}.

\subsubsection{Photoisomerization of stilbene in the presence of light-matter interaction}
\begin{figure}
  \includegraphics[width=0.44\textwidth]{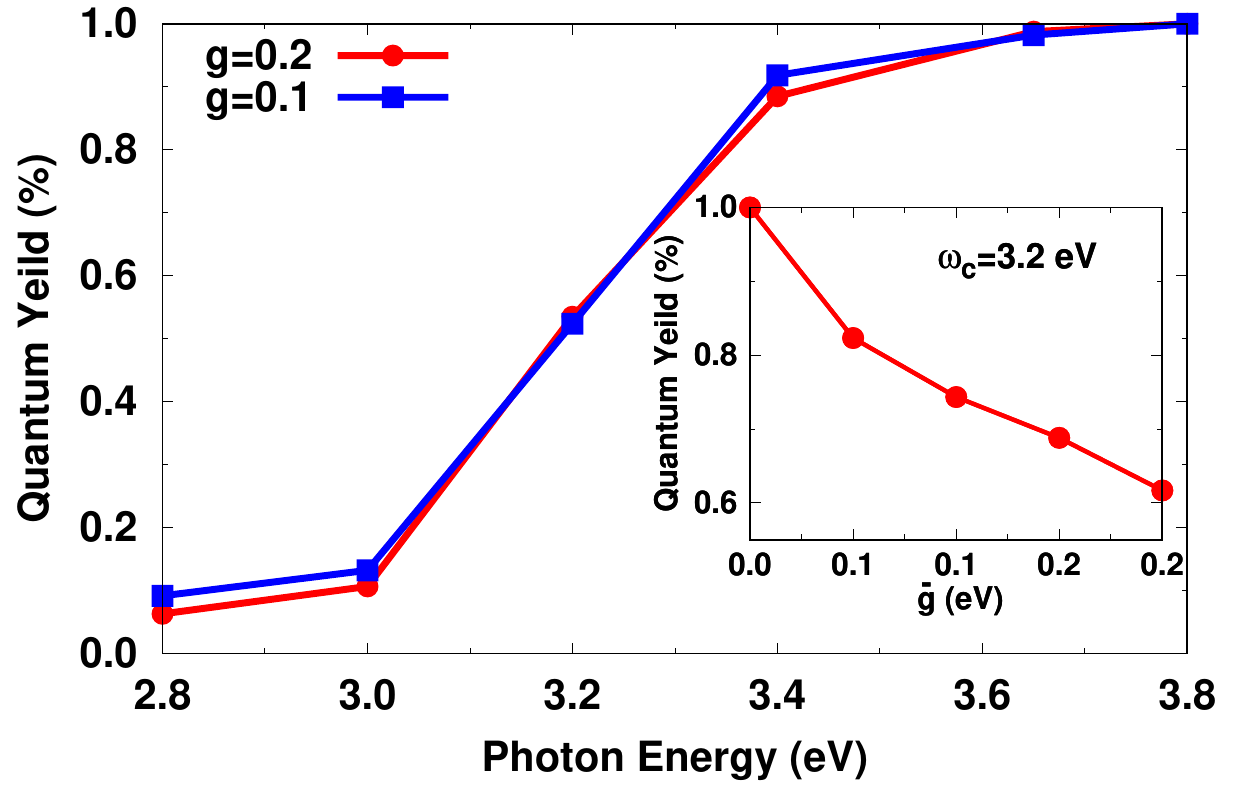}
  \caption{\label{figqy}QY as a function of photon energies and coupling strengths. QY is defined as the fraction of trajectories evolving on $S_1$ that encounter an energy gap smaller than 1.0~eV. The inset figure shows the QY as function of $\bar{g}$, which indicates stronger effect of $\bar{g}$ on QY. }
\end{figure}

When the light-matter interaction is introduced, the potential energy surfaces are altered. As shown by Eq.~\ref{cavityH}, the electronic Hamiltonian is modified by including the light-matter interaction in the cavity. In this work, a single quantized light mode, which can be microcavity modes or localized surface plasmon, is included. Without light-matter interaction, there is no interaction between the photonically excited states $|S_0,1\rangle$ and electronic excited states $|S_i,0\rangle$.
The light-matter interaction described by $\bar{g}_i(\bR)(c^\dag_i a+c_i a^\dag)$ makes the photonically excited state, $|S_0,1\rangle$ with energy  $\omega_c$, and electronic excited state, $|S_1,0\rangle$ with energy $\Omega_{1}(\bR)$, coupled.  
Consequently, two polaritonic states ($P_{L/U}$) with avoided crossing between them, are formed. As shown by  Fig.~\ref{figpes}(b-d), two polaritonic states are formed due to the hybridization between $|S_0,1\rangle$ and $|S_1,0\rangle$ states. The red and blue curves in Fig.~\ref{figpes}(b-d)  are the upper and lower polaritonic states for different photon energies. PESs of higher energy states are also plotted in Fig.~\ref{figpes2}, which indicates that hybridization between $|S_0,1\rangle$ and higher electronic states (for example $S_{2-3}$) may also exist depending on the photon energy. However, in the case of stilbene, the formed $P_{S_{2-3}}$ states do not have an obvious effect on the PES landscape and thus have a limited effect on the reactions.
The color scale represents the contribution of the two singly excited states, $|S_1,0\rangle$ and $|S_0,1\rangle$, to the polaritonic states. The color scale of Fig.~\ref{figpes}(b-d) implies that the light-matter hybridization is more significant at the points where the energies of $|S_0,1\rangle$ and $|S_1,0\rangle$ are close, i.e., the molecular detuning $|\Delta(\bR)|\equiv |\omega_c-\Omega_{1}(\bR)|$ is comparable to or smaller than $\bar{g}_1(\bR)|$. At the points that $|\Delta| \gg |\bar{g}_1(\bR)|$, the polaritonic states are almost the same as the bare ones. Especially, when the photon energy is higher than the $S_1-S_0$ gap, the $S_1$ state is not affected by the light-matter interaction. Consequently, the dynamics on the $S_1$ PES surface is not affected by the light-matter interaction, and the photoisomerization pathway and QY are the same as they would be in the absence of light-matter interaction as shown by Fig.~\ref{figqy}.

When the photon energy becomes equal or smaller than the $S_1-S_0$ gap (3.65~eV in average), the interaction between the $|S_0,1\rangle$ and $|S_1,0\rangle$ states splits at the crossing. Fig.~\ref{fig2}(c) plots the absorption spectrum of the \textit{cis} isomer with and without the light-matter interaction, where $g=0.1$ (in unit of eV/Bohr) and photon energy is 3.65~eV in resonance with the average energy of $S_1$ states. The absorption spectrum clearly shows the splitting of $S_1$ state into two bright polaritonic states $P_{L/U}$ after the light-matter interaction is introduced. In this case, the distance between the two polaritonic peaks, shown by the red line of Fig.~\ref{fig2}(c), is the Rabi splitting defined by Eq.~\ref{rabi1}, i.e., $2 \bar{g}_1(\bR) \simeq 0.2$~eV, and the PESs of the two polaritonic states are different from the that of bare ones ($S_0$ and $S_1$ states). The obvious Rabi splitting between the two polaritonic peaks also indicates the strong coupling is achieved. But it should be noted that the $P_U$ peak also has contribution from the $S_{2-3}$ states (the shoulder next to the $S_1$ state, $\sim$3.8~eV, as shown by the black line of Fig.~\ref{fig2}(c)). Even though $S_{2-3}$ states are also coupled to the cavity, the effect of strong coupling on these states is smaller as it is off-resonance with the cavity mode and the transition dipole is smaller.  Consequently, $S_{2-3}$ peaks are slightly shifted and overlap with the upper polaritonic peak, resulting in the $P_U$ peak (the peak is not pure $P_U$ state) shown by the red line of Fig.~\ref{fig2}(c). In regions where two coupled PESs are in resonance, the shape of the polaritonic PES exhibits mixed characteristics of each state due to strong coupling. Since the PES of photonically excited state $|S_0,1\rangle$ is a copy of ground state PES, hybridization leads to the appearance a local minimum in the lower polaritonic PES ($P_L$). Consequently, a potential energy barrier for isomerization can develop on the $P_L$ surface. Following excitation to the $P_U$ states, the dynamics may involve the transition through the crossing to the $P_L$ states. After this transition, the wavepacket continues to propagate towards the CI and trigger the photoisomerization. However, there is also a possibility that the nuclear dynamics can return back to the \textit{cis} configuration driven by the gradient of $P_L$ surface as illustrated by the purple arrow in Fig.~\ref{figpes}(d). As a result, the QY of \textit{cis-trans} isomerization can be suppressed. As confirmed by Figs.~\ref{figfrac} and~\ref{figqy}, the QY of isomerization is suppressed when the photon energy is resonant with the $S_1-S_0$ gap.

Because the photonically excited state $|S_0,1\rangle$ is controllable via photon energy, the landscape of the polaritonic PESs can be tuned via the photon energy and light-matter coupling strength as shown in Fig.~\ref{figpes}(b-d). Consequently, the non-adiabatic dynamics on the $P_{L/U}$ states, as well as QY, are dependent on photon energy and coupling strength. For smaller photon energies, the minimum on the lower polaritonic PES becomes deeper, confirmed in Figs.~\ref{figpes}(b-d). Consequently, there is a chance that the wavepacket will be trapped in the minimum and isomerization will be suppressed. As shown by the Fig.~\ref{figfrac}, the fraction of trajectories with $\Delta E =\Omega_{1}(\bR)<1.0$~eV is significantly altered by different photon energies. In general, lower photon energy decreases the fraction of trajectories that can encounter an energy gap smaller than 1.0~eV suppressing isomerization. In addition, the isomerization is also slightly affected by different coupling strengths. Larger light-matter coupling strength results in increased splitting at the crossing as shown by Eq.~\ref{rabi1}, owing to a deeper landscape on $P_L$ PES toward the \textit{cis}-configuration, as shown by the comparison between the blue and black curves in Fig.~\ref{figcrossing}. Consequently, the wavepacket that passes through the crossing acquires slightly larger probability of getting trapped in the minimum. Hence, photoisomerization is slightly suppressed by a large coupling strength. But this effect is screened by the large fluctuation of transition dipole, as discussed later. Besides, since the photon energy has larger effect in developing deeper minima in $P_L$ PES, the effect of light-matter coupling strength on isomerization is less significant compared to that of photon energy. However, it should be noted that the strong light-matter coupling occurs in the non-equilibrium regions of the $S_0$ surface when the photon energy is smaller than the $S_1-S_0$ gap (3.65~eV) of the equilibrium configurations. The condition can only be achieved experimentally by employing plasmonic cavities in which strong light-matter interaction at the single-molecule level can be achieved. In contrast, in photonic cavities, a large concentration of molecules is essential to reach strong coupling regime. In this case, it is difficult to maintain the concentration of molecules, therefore the strong coupling,  in certain non-equilibrium regions.

In our simulation, the initial population on the upper polaritonic state is initiated by an instantaneous transition. Single different initial geometries and velocities are used, all vibrational states that are accessible from the ground state are excited in the trajectory swarm. If the wavepacket is trapped by the local minimum of the $P_L$ surface without undergoing the isomerization process, the final state of the molecules will be determined by other effects. The excited wavepacket will dissipate its energy to the vibrational degrees of freedom on a timescale of picoseconds. In addition, the excited state wavepacket can decay radiatively back to the ground state. 

Fig.~\ref{figqy} shows the calculated QY as a function of photon energies and light-matter coupling strength for our model stilbene system. When the photon energy is larger than a certain value (3.8~eV being slightly larger than the $S_1$-$S_0$ energy gap of \textit{cis}-configuration), the QY is not affected by the light-matter interaction since the PES of $P_{L/U}$ states are the bare $S_{1/0}$ states as argued above. With decreasing photon energy, the PES of $P_L$ establishes a deeper minimum. The NAC between $P_U$ and $P_L$ states induced by vibrations initiates the transition through the crossing, resulting in the occupation of the $P_L$ state. Consequently, for a photon with lower energy, the wavepacket initially excited on $P_U$ state can easily get trapped by the local minimum of $P_{L}$ after transition through the avoided crossing guided by the polaritonic NACs. The numerical simulations shown in Fig.~\ref{figfrac} and Fig.~\ref{figqy} indicate that the QY of photoisomerization is significantly suppressed for lower photon energy because the wavepacket is trapped by the lower polaritonic state. Besides, one may expect that the effect of light-matter interaction will be minimized at the limit of $\omega_c\rightarrow 0$ because the PESs of the polaritonic states should be the same as that of the bare molecule at this limit. Fig.~\ref{fig-e} plots the QY with smaller photon energies. Even though the RWA is invalid when $\omega_c$ is far from resonance, the results qualitatively demonstrate the recovery of the bare NAMD results when $\omega_c\rightarrow 0$.
In contrast, as argued before, the suppression of the photoisomerization is less sensitive to the coupling strength, which can be attributed to the large fluctuation of the transition dipole moment. The light-matter coupling strength is dependent on the transition dipole ($\bar{g}_i(\bR)=g\mu_i(\bR)$) which varies with geometries. Such dependence has also been reported in previous works~\cite{Fregoni2018,jcp4941053}. Since the transition dipole can change dramatically during the dynamics, the splitting between the polaritonic PESs can fluctuate between very small and large, even at the same reaction coordinate $\theta$ (since other coordinates can be different). For comparison, a set of simulations with $\bar{g}_i(\bR)$ fixed as a constant was performed, as shown in the inset of Fig.~\ref{figqy}. The results indicate that the light-matter coupling strength has a more significant effect once it can be fixed as a constant value instead of fluctuating with transition dipole. The comparison demonstrates the fluctuation in the transition dipole screens the effect of coupling strength, which indicates that it is harder to tune the photochemistry via the coupling strength compared to the photon energy.

\begin{figure}
  \includegraphics[width=0.44\textwidth]{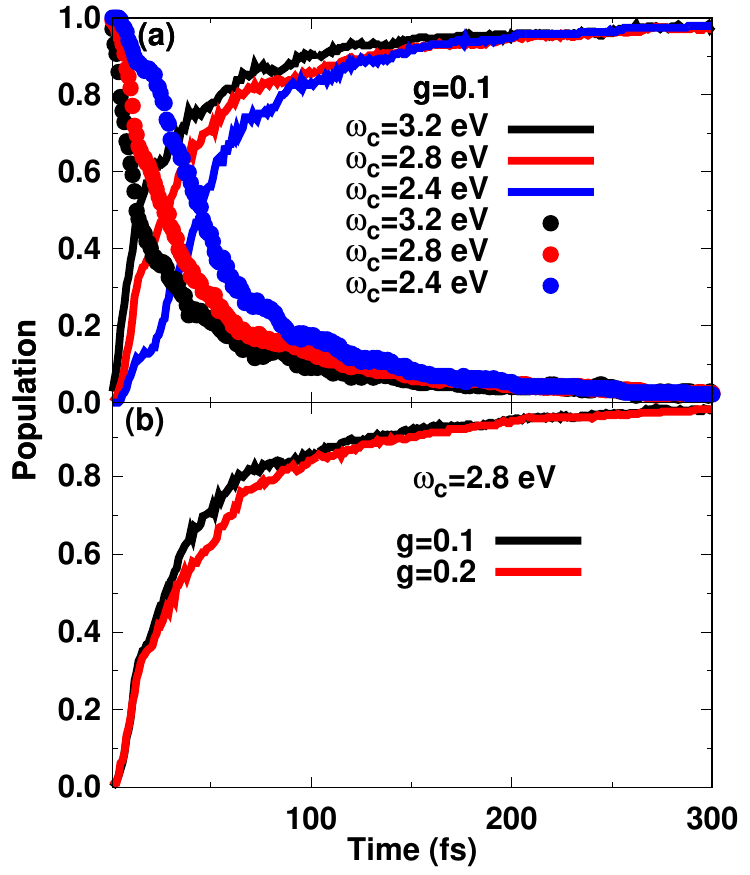}
  \caption{\label{figpop}Populations on $P_L$ (solid lines) and $P_U$ (dots) as a function of time for different (a) photon energies and (b) coupling strength. The results indicate the relaxation processes is altered by different photon energies and coupling strengths.}
\end{figure}

In addition, our simulations indicate that the upper polaritonic PES can undergo efficient relaxation to the lower polaritonic PES, which is consistent with previous works~\cite{PRB84205214}. Fig.~\ref{figpop} plots the populations of the two polaritonic states as a function of time following photoexcitation to the initial state $P_U$. After the photoexcitation, the molecule relaxes nonadiabatically to the $P_L$ state. The nonadiabatic relaxation to the lower polaritonic state takes place quickly within 100~fs. Since the PESs of $P_{L/U}$ states are altered by different photon energies, the relaxation process is also affected. As shown by the purple arrow of Fig.~\ref{figpes}, the relaxation pathway becomes longer when a smaller photon energy is imposed. In general, smaller photon energy makes the crossing points between the $P_{L/U}$ states farther from \textit{cis} configurations. Consequently, population relaxation from the $P_U$ state to the minimum of $P_L$ takes slightly longer time as shown by Fig.~\ref{figpop}. Moreover, the comparison between the isomerization and population dynamics shown by Fig.~\ref{figfrac} and Fig~\ref{figpop} indicates the \textit{cis-trans} isomerization would occur on a much longer timescale after relaxation to the $P_L$ state~\cite{jpca8b09103}. Since Eq.\ref{rabi1} indicates that the Rabi splitting is proportional to the light-matter coupling strength, Rabi splitting at the avoided crossings becomes larger with increasing coupling strengths. As a result, as shown in Fig.~\ref{figpop}, the population dynamics is also slightly slowed down by a larger $g$ due to the larger gap at the avoided crossing. But it should be noted that relaxation dynamics strongly depends on the PES landscape and CI locations, all of which are system-dependent.

\section{Discussions and summary}
\label{summary}
Recent experiments on \textit{polaritonic chemistry} demonstrated the potential applications of strong light-matter interaction in manipulation of the PES landscapes and corresponding chemical reactions at the molecular level~\cite{accounts6b00295}. This experimental advance requires the development of concomitant theories and efficient computational methods for modeling the dynamics of molecules in the presence of strong light-matter interactions. In principle, modeling of polaritonic chemistry is nontrivial, which requires the treatment of electrons, nuclei and photons, on equal footing~\cite{Flick3026} in the regime of non-adiabatic dynamics. However,  the full non-adiabatic quantum treatment of electrons and nuclei is already computationally expensive and is limited to a few degrees of freedom~\cite{cr7b00423}. Simulation of realistic molecules calls for the need of development of reasonable approximations cutting computational cost of atomistic simulations. The MQC method is one example of the efficient method for modeling the NAMD of isolated molecules. 

In this work, we have extended the capability of NAMD to handle the nuclear dynamics on hybrid light-matter potential energy surfaces of realistic molecules interacting strongly with the quantized photon in the cavity. TDSCF method was employed to calculated the excited state properties of the isolated molecules. The RWA was then adopted for the calculation of polaritonic energies and WFs. Combining the gradients of isolated molecules and polaritonic WFs, the forces on polaritonic PESs and NAC between different polaritonic states are analytically calculated, which allows us to simulate the NAMD in the presence of light-matter interaction on-the-fly. The resulting NAMD formalisms, including Mean-field Ehrenfest dynamics and the FSSH algorithms, are almost the same as their respective counterparts for bare molecules except the energies, forces, and NAC are dressed by the light-matter interactions. The FSSH scheme of NAMD for strong light-matter interaction is implemented in the semiempirical NEXMD code~\cite{ar400263p}. Our work thus provides an efficient approximate framework for the modeling of polaritonic chemistry in realistic molecular systems. The proposed methodology is exemplified for \textit{cis-trans} isomerization of stilbene to demonstrate its utility in controlling photochemistry. 

Our NAMD simulations of photoisomerization in stilbene have shown that the strong coupling of a single molecule to a confined photon mode can strongly affect the PES landscape and therefore alter the photochemical processes. Our numerical simulations indicate that, by tuning the cavity properties including photon energies and light-matter coupling strengths, the \textit{cis-trans} reaction can be suppressed. It should be noted that our result for stilbene isomerization without cavity is consistent with experimental findings~\cite{cr00003a007}. Furthermore, our result for isomerization with polaritonic states is consistent with the one-dimensional quantum non-adiabatic dynamics simulations~\cite{Galego2016}. This agreement provides additional validation of our MQC method for the modeling of polaritonic chemistry. 
 Moreover, our NAMD simulations demonstrate that the large fluctuation of transition dipole during the reaction screens the effect of the light-matter coupling strength, making it harder to tune the photochemical probability using the coupling strength compared to the photon energy.
Moreover, the non-adiabatic relaxation rates are also altered due to the fact that the relaxation pathway is modified by the polaritonic PESs. Our stilbene testbed system and simulations were deliberately chosen to demonstrate control of photochemistry through modifications of molecular PESs due to strong light-matter interactions. Additional tunability is possible through variability of NAC values, modification of reaction kinetics, modulation of photoexcited pathways, coherent phenomena~\cite{Nelson2018}, etc. Here such a tool as NEXMD, providing detailed atomistic information, is particularly helpful for predicting and exploring conceivable physical processes. This methodology will benefit the optimization and control of photophysical and photochemical properties of molecular systems.

The recent achievement of strong coupling at the single-molecule level implies the single-molecule manipulation~\cite{RN638}. On the other hand, many experiments on strong coupling with organic molecules have explored the collective behaviors in which $M$ molecules coherently interact with a single photon mode. According to the TC model, $M$ identical molecules interacting with a single mode can lead to an enhancement of the Rabi frequency by a factor of $\sqrt{M}$. The collective behavior is not described in this work but it can be readily simulated by including more than one molecule in the simulation~\cite{jctc7b00388}. However, it should be noticed that $M$ molecules cannot be exactly identical in realistic conditions due to the thermal fluctuation. For $M$ identical molecules, many interesting features have been found in previous works, including single excitation induced collective photochemical reactions~\cite{PRL119136001}. But the effect of the thermal fluctuation in a realistic $M$-molecule ensemble may reduce the collective behavior to some extent, which will be explored in future works. Furthermore, the plasmonic cavity suffers from strong dissipations. As a result, the polaritonic states are also subject to significant dissipations. The dissipation of polaritonic states may limit the applications in photochemistry. In the framework of open quantum systems, the dissipation can be taken into account by introducing additional dissipative terms (standard treatment in open quantum systems)~\cite{BRE02} to Eqs.~\ref{ceom} or \ref{fssheom} to account for the photon losses in the cavity. Alternatively, the dissipation can also be introduced via quantum jump~\cite{rmp70101} or a Monte Carlo approach~\cite{Fregoni2018}. In addition, when the light-matter coupling enters the ultrastrong regime, the solution of the full Hamiltonian may require large Hilbert space to converge~\cite{jpclett6b00864}. Thus, a general and efficient treatment of polaritonic energies beyond RWA is essential for polaritonic chemistry.

\begin{acknowledgments}
The work at Los Alamos National Laboratory (LANL) was supported by the LANL Directed Research and Development Funds (LDRD) and performed in part at the Center for Integrated Nanotechnologies (CINT), a U.S. Department of Energy, Office of Science user facility at LANL. This research used resources provided by the LANL Institutional Computing (IC) Program. LANL is operated by Triad National Security, LLC, for the National Nuclear Security Administration of the U.S. Department of Energy (Contract No. 89233218NCA000001). 
\end{acknowledgments}

\input{appendix}

\bibliography{ref}

\end{document}

%% file: appendix.tex

{\centering
\Large 
\begin{center}
\textbf{APPENDIX}
\end{center}
}

\setcounter{section}{0}
\setcounter{equation}{0}

\renewcommand{\theequation}{A\arabic{equation}}

\section{Derivative of transition dipole}\label{appsec2}
The transition dipole between the ground state and $i^{th}$ excited state is~\cite{qua1528}
\begin{equation}\label{eqdipole}
\mu_i(\bR) = \sum_{\mu\nu} \langle \mu|\bR|\nu\rangle \xi_{i,\nu\mu}
\end{equation}
where $\xi^i$ is the transition density matrix~\cite{qua1528}. The transition density matrix and excited
state energies $\Omega_i$ are calculated from the Liouville equation
\begin{equation}
\mathcal{L}\xi_i=\Omega_i\xi_i
\end{equation}
where the action of Liouville operator on $\xi_i$ is~\cite{jcp4905828}
\begin{equation}\label{liouville}
\mathcal{L}\xi_i=[h(\rho_0),\xi_i]+[V(\xi_i),\rho_0]
\end{equation}
where $h(\rho_0)=t+V(\rho_0)$ is the Fock matrix and $V$ the Coulomb-exchange electronic operator, and $\rho_0$ is the ground state density matrix.

According to Eq.~\ref{eqdipole}, the derivative of the transition dipole reads
\begin{align}
\frac{\partial\mu_i}{\partial \bR_a}= & \sum_{\mu\nu}\left[\frac{\partial\langle \mu| \bR|\nu\rangle}{\partial \bR_a}
\xi_{i,\nu\mu}+\langle \mu|\bR|\nu\rangle \frac{\partial \xi_{i,\nu\mu}}{\partial\bR_a}\right]
\end{align}
Following the derivation in Ref.\citenum{qua1528}, the derivative of transition density is obtained by applying $\partial/\partial x$ to Eq.~\ref{liouville}, i.e., 
\begin{equation}\label{eqdxi}
\left(\mathcal{L}-\Omega_i\right)\xi^x_i = -(\mathcal{L}^x-\Omega^x_i)\xi_i.
\end{equation}
where $\xi^x_i$ denotes $\frac{\partial \xi_i}{\partial x}$. The right-hand side of above equation is assumed to be known and will be denoted $C^x_i$ which can be directly computed. According to the definition of $\mathcal{L}$ operator, $\mathcal{L}^x \xi_i$ can be rewritten as
\[
\mathcal{L}^x\xi_i= [h^x,\xi_i]+[V(\rho^x_0),\xi_i]+ [V^x(\xi_i),\rho_0]  +[V(\xi_i),\rho^x_0] 
\]
Hence, $C^x_i$ can be rewritten as 
\begin{align}
C^x_i=&-[h^x,\xi_i]-[V^x(\xi_i),\rho_{0}]-[V(\rho^x_{0}),\xi_i]
\nonumber\\ &-[V(\xi_i),\rho^x_0]+\Omega^x_i\xi_i.
\end{align}
The analytical derivatives of Fock matrix $h^x$, Coulomb-exchange matrix $V^x$, and bare excited-state energies $\Omega^x_i$ can be easily calculated in TDSCF methods~\cite{ar400263p,C3CP51514A}. Since the ground state density matrix commutes with Fock matrix, i.e., $[h(\rho_0),\rho_0]=0$, the $\rho^x_0$ can be readily calculated from
\begin{equation}
[h(\rho_0),\rho^x_0]=-[h^x,\rho_0]
\end{equation}
by employing the iterative Bi-Conjugate Gradient Stability (BICGStab) method~\cite{siam913035}. After $\rho^x_0$ is calculated, $C^x_i$ is known. Then $\xi^x_i$ can be calculated from the Eq.~\ref{eqdxi} by the BICGStab method. 

\section{Derivation of NAC}
\label{appsec1}
\subsection{NAC without light-matter interaction}
The definitions of non-adiabatic coupling vector (NACR) and time-dependent non-adiabatic coupling (NACT) are~\cite{jcp459170}
\begin{align}
\bd_{kl}=&\langle \psi_k(\bR)|\nabla_\bR\psi_l(\bR)\rangle
\nonumber\\
\dot{\mathbf{R}}\cdot \bd_{kl}=&\langle\psi_k(t)|\nabla_t\psi_l(t)\rangle, k\neq l
\end{align}
According to the definition of NAC (either NACR or NACT), the derivative of the WF with respect to time or the nuclear coordinate is required. Since NACT and NACR have similar structure, their derivations share the same procedure. We denote the time or nuclear coordinates as $x$, then we can calculate the NAC by perturbing the molecular Hamiltonian by introducing a small displacement $\Delta$ for a given coordinate $x$. According to the perturbation theory, the perturbed Hamiltonian can be expressed in the first order as
\begin{equation}
\hat{H}(x+\Delta)=\hat{H}(x)+\frac{\partial \hat{H}}{\partial x}\Delta+o(\Delta^2).
\end{equation}
Consequently, the standard perturbation theory can be applied to get the first-order approximations to the eigenvalues $E_k(x+\Delta)$ and eigenvectors $|\psi_k(x+\Delta)\rangle$ of the perturbed Hamiltonian $H(x+\Delta)$.
\begin{equation}
E_k(x+\Delta)=E_k(x)+\langle \psi_k(x)|\frac{\partial \hat{H}}{\partial x}|\psi_k(x)\rangle\Delta + o(\Delta^2).
\end{equation}
This can also be immediately used to calculate the derivative of the eigenvalue itself
\begin{equation}
\frac{\partial E_k(x)}{\partial x}=\lim_{\Delta\rightarrow 0}\frac{E_k(x+\Delta)-E(x)}{\Delta}
=\langle \psi_k(x)|\frac{\partial \hat{H}}{\partial x}|\psi_k(x)\rangle,
\end{equation}
which is consistent with Hellmann-Feynman theorem~\cite{PR56340}. The same argument can be used to calculate the derivative of the WF as well. The first-order correction to the WF is
\begin{equation}
|\psi_k(x+\Delta)\rangle = |\psi_k(x)\rangle +\Delta \sum_{j>0;j\neq k}
|\psi_j(x)\rangle \frac{\langle \psi_j(x)|\partial \hat{H}/\partial x|\psi_k(x)\rangle}{E_k(x)-E_j(x)}.
\end{equation}
Consequently, the derivative of WF can be written as 
\begin{align}
\frac{\partial |\psi_k(x)\rangle}{\partial x}= &\lim_{\Delta\rightarrow 0}\frac{|\psi_k(x+\Delta)\rangle-|\psi_k(x)\rangle}{\Delta}
\nonumber\\
=&\sum_{j\neq k} |\psi_j(x)\rangle\frac{\langle \psi_j(x)|\partial \hat{H}/\partial x|\psi_k(x)\rangle}{E_k(x)-E_j(x)} .
\end{align}
Therefore, the NAC is~\cite{jcp480511}
\begin{align}
\langle\psi_k(x)|\nabla_x\psi_l(x)\rangle &=\langle \psi_k(x)|\frac{\partial}{\partial x}|\psi_l(x)\rangle
\nonumber\\ &=  \langle \psi_k(x) | \sum_{j\neq l} |\psi_j(x)\rangle\frac{\langle \psi_j(x)|\partial \hat{H}/\partial x|\psi_l(x)\rangle}{E_l(x)-E_j(x)}
\nonumber\\ &=  \frac{\langle \psi_k(x)|\partial \hat{H}/\partial x|\psi_l(x)\rangle}{E_l(x)-E_k(x)}, k\neq l.
\end{align}
Introducing a basis set, with the corresponding Fermionic creation and annihilation operators $c^\dag_n$ and $c_n$, the derivative of the molecular Hamiltonian can be written in terms of the creation and annihilation operators, $\frac{\partial \hat{H}}{\partial x}=\sum_{\mu\nu} h^x_{\mu\nu}c^\dag_\mu c_\nu$. Thus,  the NACT can be rewritten as~\cite{jp109522g}
\begin{equation}
A_{kl}\equiv\dot{\mathbf{R}}\cdot \bd_{kl}= \frac{\text{Tr}[\rho_{kl}h^t]}{E_k(t)-E_l(t)}, k\neq l,
\end{equation}
where $h^t$ is the time derivative of Fock matrix calculated by 
\begin{equation}
h^t=\lim_{\Delta\rightarrow 0}\frac{h(t+\Delta t/2)-h(t-\Delta t/2)}{\Delta t},
\end{equation}
and $(\rho_{ik})_{\mu\nu}=\langle \psi_i|c^\dag_\mu c_\nu|\psi_k\rangle$ denotes the one-electron transition density matrix. The time-dependent Hatree-Fock (TDHF) method is employed to calculate the matrix element of the one-electron transition density matrix~\cite{jcp480443}. The expression of NAC involving ground state is even simpler and only require the knowledge of $(\rho_{i0})_{\mu\nu}=\langle i|c^\dag_\mu c_\nu|0\rangle=(\xi_i)_{\mu\nu}$ and $(\rho_{0i})_{\mu\nu}=\langle 0|c^\dag_\mu c_\nu|i\rangle=(\xi^\dag_i)_{\mu\nu}$~\cite{qua1528}. Consequently,
\begin{equation}
A_{i0}= \frac{\text{Tr}[\xi_i      h^t]}{E_i},\quad\quad 
A_{0i}=-\frac{\text{Tr}[\xi^\dag_i h^t]}{E_i}, 
\end{equation}
The matrices $\{\xi_i\}$ are obtained from the CIS technique~\cite{Mukamel781}. The transition density matrices between excited states, $\rho_{ik}$, can also be expressed in terms of the CIS. 

\subsection{NAC with strong light-matter interaction}
When strong light-matter interaction is present, the excited states and corresponding WFs are altered. Consequently, the NAC is modified by the strong light matter interaction. The NAC with strong light-matter interaction is 
\begin{equation}\label{eq:akl2}
\bd_{KL}=\langle \Psi_K| \nabla_\bR|\Psi_L\rangle
=\frac{\langle \Psi_K(t)|\hat{H}^\bR|\Psi_L(t)\rangle}{E_L(t)-E_K(t)}.
\end{equation}
Because the polaritonic states are related to the bare states as
\begin{equation}\label{eq:k2}
|\Psi_K\rangle=\sum^{N}_{i=0} \beta^K_j |\psi_j\rangle
\end{equation}
The NACT between polaritonic states is
\begin{align}\label{eq:klht2}
\langle \Psi_K|\hat{H}^\bR|\Psi_L\rangle = & \left( \sum^{N}_{i=0} \beta^{*,K}_i \langle \psi_i| \right) 
H^\bR |\left( \sum^{N}_{j=0} \beta^{L}_j | \psi_j\rangle \right) 
\nonumber\\=&
\sum^N_{i,j=0}\beta^{*,K}_i\beta^L_j \text{Tr}[\rho_{ij}h^\bR]
\end{align}
Substituting the above equation, Eq.~\ref{eq:klht2}, into Eq.\ref{eq:akl2} gives the NAC for strong light-matter interaction. 

\begin{figure}[H]
  \includegraphics[width=0.42\textwidth]{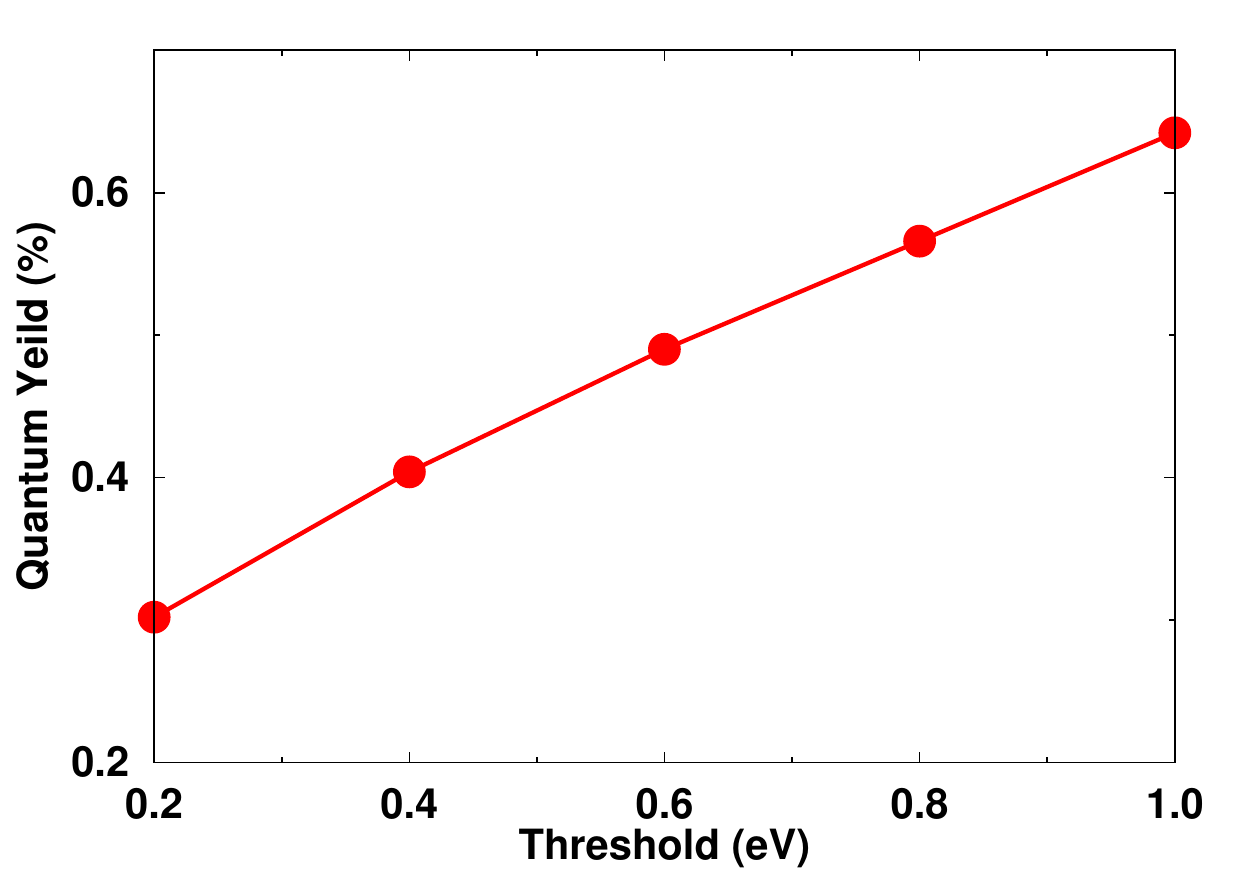}
  \caption{\label{fig-delta}Quantum yield as function of the threshold. $\omega_c=3.2$~eV.}
\end{figure}

\begin{figure}[H]
  \includegraphics[width=0.40\textwidth]{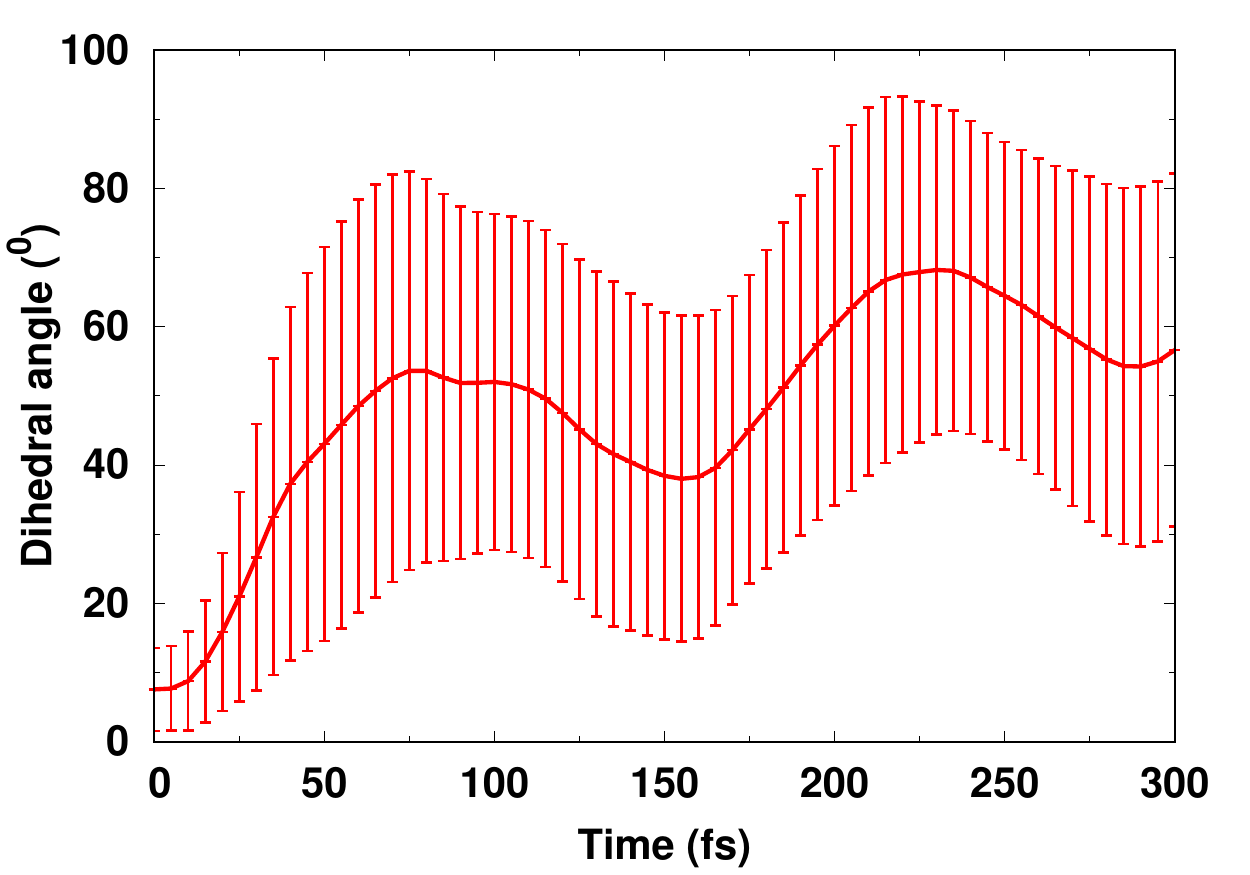}
  \caption{\label{fig-dihedral}
Time evolution of mean dihedral angle (with error bars) in absence of light-matter interaction.}
\end{figure}

\vskip -10pt
 
\begin{figure}[H]
  \includegraphics[width=0.42\textwidth]{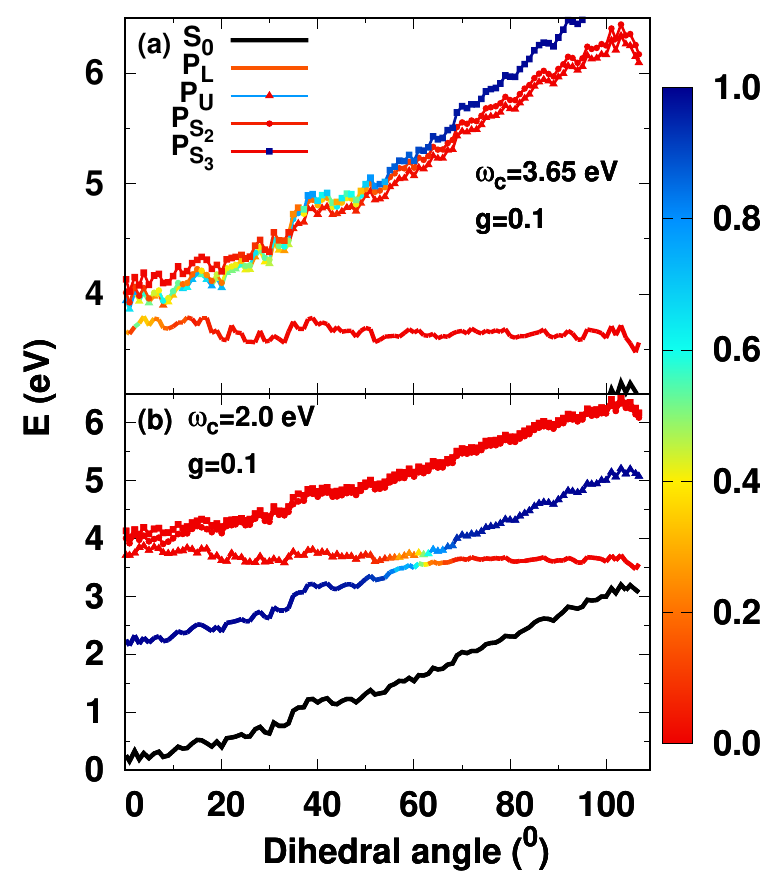}
  \caption{\label{figpes2}PESs of \textit{cis} isomer with light-matter interaction with higher electronic excited state ($S_{2-3}$): a) $\omega=3.65$~eV, b) $\omega=2$~eV.}
\end{figure}

\vskip -10pt

\begin{figure}[H]
  \includegraphics[width=0.40\textwidth]{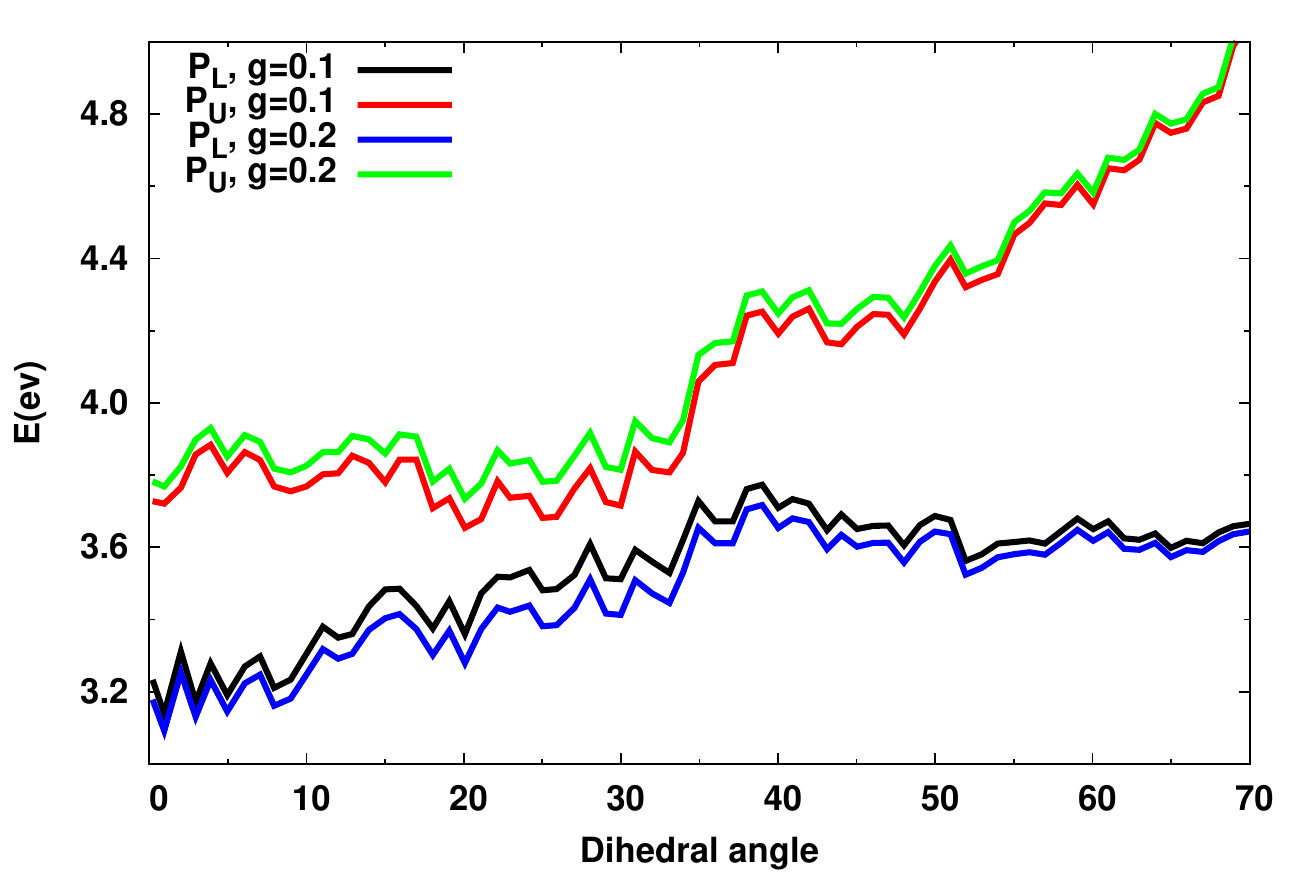}
  \caption{\label{figcrossing}Avoided crossings between two polaritonic states for different coupling strength.}
\end{figure}

\vskip -10pt

\begin{figure}[H]
  \includegraphics[width=0.42\textwidth]{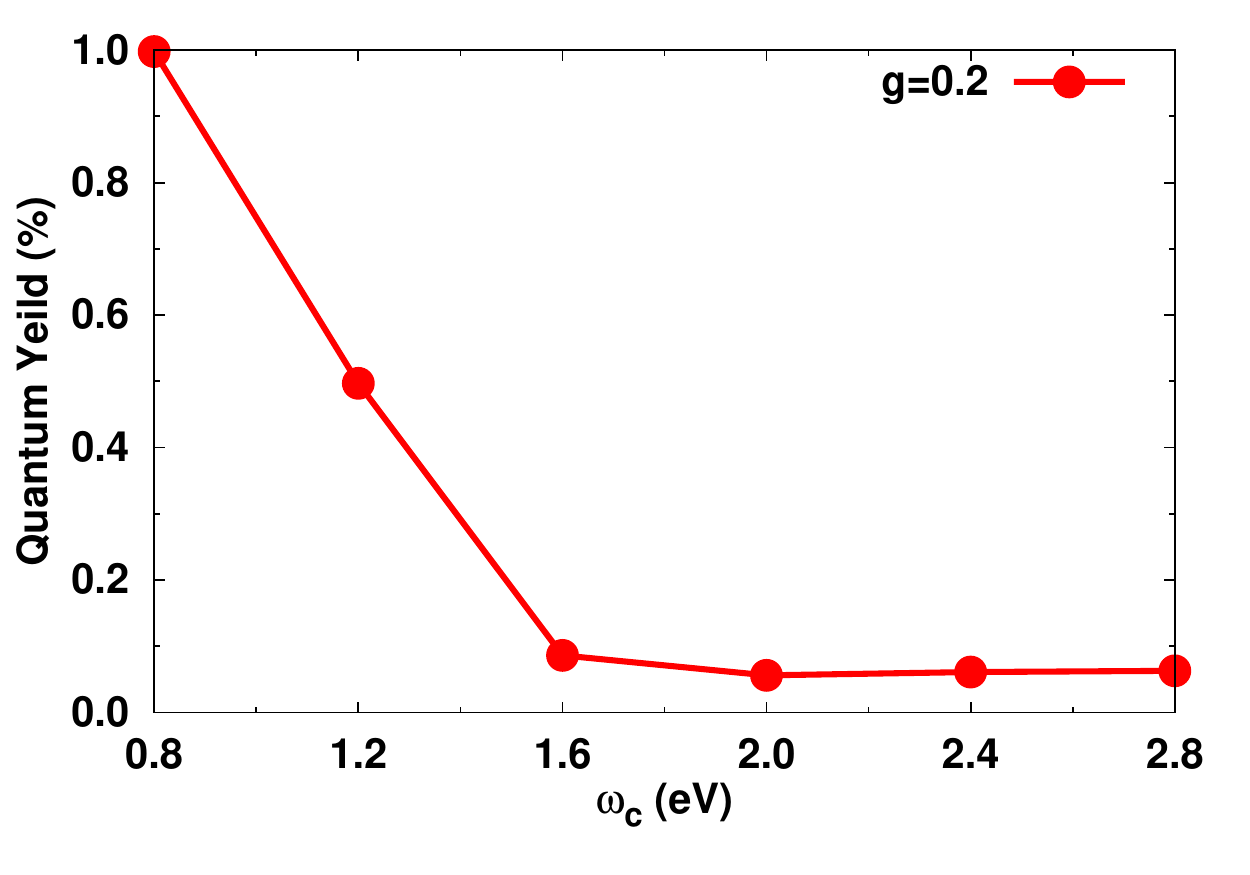}
  \caption{\label{fig-e}Quantum yield as function of photon energies for $\omega_c\leq 2.8$~eV.}
\end{figure}